\documentclass{article}

\usepackage{amsmath,amssymb,amsthm,amscd,euscript,amsfonts}
\usepackage{mathtools}
\usepackage{color}
\usepackage[pdftex]{graphicx}
\usepackage[toc,page]{appendix}
\usepackage[utf8]{inputenc}
\usepackage[nottoc]{tocbibind}
\usepackage{authblk}
\DeclareGraphicsExtensions{pdf, jpg, jpeg, png}

\newcommand{\nid}{\noindent}

\renewcommand{\o}{\otimes}
\newcommand{\eps}{\varepsilon}

\input xy
\usepackage[color,matrix,arrow]{xy}
\xyoption{all}

\def \c{\mathbb C}
\def \l{\mathcal L}

\newtheorem{theorem}{Theorem}[section]

\newtheorem{remark}[theorem]{Remark}

\title{On the boundaries of quantum integrability for the spin-$1/2$ Richardson--Gaudin system}
\author{Inna Lukyanenko, Phillip Isaac, Jon Links}
\affil{\textit{Centre for Mathematical Physics}, \\
\textit{School of Mathematics and Physics}, \\
\textit{The University of Queensland, 4072, Australia}}

\date{}

\begin{document}
\input epsf

\maketitle
\vspace{-1.1cm}
\begin{abstract}
We discuss a generalised version of Sklyanin's Boundary Quantum Inverse Scattering Method applied to the spin-$1/2$, trigonometric $sl(2)$ case, for which both the twisted-periodic and boundary constructions are obtained as limiting cases. We then investigate the quasi-classical limit of this approach leading to a set of mutually commuting conserved operators which we refer to as the trigonometric, spin-$1/2$ Richardson--Gaudin system. We prove that the rational limit of the set of conserved operators for the trigonometric system is equivalent, through a change of variables, rescaling, and a basis transformation, to the original set of trigonometric conserved operators.  Moreover we prove that the twisted-periodic and boundary constructions are equivalent in the trigonometric case, but not in the rational limit. 
\end{abstract}
\vspace{-0.4cm}
\tableofcontents

\section{Introduction}

In 1988 Sklyanin proposed the Boundary Quantum Inverse Scattering Method \cite{s88}. Based on the Yang--Baxter Equation \cite{b72,m64,y67} and the reflection equations \cite{c84}, this formalism permits the construction of one-dimensional quantum systems with integrable boundary conditions, and the derivation of associated exact Bethe Ansatz solutions. The examples of the $XXZ$ and $XYZ$ spin chains, the non-linear Schr\"odinger equation, and the Toda chain are discussed in \cite{s88}. The method has been widely applied for the construction and analyses of one-dimensional quantum models with integrable boundaries, and related mathematical structures, for more than two decades, e.g. \cite{aacdfr03,bc13,cysw13,dlr13,fk10,fk93,fs99,g08,ik14,iwwz13,ks92,ml06,mn92,mrm05,n12}. 

In more recent times integrable models based on the {\it quasi-classical} limit of the Yang--Baxter Equation (also known simply as the {\it classical} Yang--Baxter Equation) have come to more prominence, in some part due to connections with pairing Hamiltonians applied to studies of superconductivity. This direction of research was motivated by experiments conducted on metallic nanograins in the 1990s, reviewed in \cite{vdr01}, and the re-examination of Richardson's hitherto little-known exact solution of the $s$-wave pairing Hamiltonian from 1963 \cite{r63}. Richardson's approach is akin to the co-ordinate Bethe Ansatz that Bethe adopted for deriving the solution of the $XXX$ chain \cite{b31}, which does not rely on a solution of the Yang--Baxter Equation.   
Also without utilising the Yang--Baxter Equation, Gaudin  provided a general algebraic formulation for constructing integrable systems related to the $sl(2)$ Lie algebra \cite{g76}. In doing so he  obtained the exact solution for a class of interacting spin models, and the Dicke Hamiltonian. These have a similar form of Bethe Ansatz Equations as those of Richardson's solution. It has become commonplace to refer to models obtained through representations of this algebra, including higher spin versions, as Richardson--Gaudin systems. Independent of knowledge of the works by Richardson and Gaudin, in 1997 Cambiaggio, Rivas and Saraceno determined a set of conserved operators for the $s$-wave pairing Hamiltonian \cite{crs97}. The eigenvalues of the conserved operators were obtained by Sierra using conformal field theory methods \cite{s00}. Gaudin's algebra admits elliptic, trigonometric, and rational function parametrisations. Later work established that Richardson's solution could be derived through a representation of Gaudin's algebra for the rational parametrisation, and generalisations could be obtained in the trigonometric case \cite{ado01,des01}\footnote{The elliptic case is generally not considered. It breaks $u(1)$ symmetry leading to non-conservation of particle number.}.
  
The works of Richardson and Gaudin provided examples of Bethe Ansatz solutions for integrable systems in the quasi-classical limit {\it avant la lettre}. It has since been clarified that Richardson's solution for the $s$-wave model, and the conserved operators, may be obtained as the quasi-classical limit of the twisted-periodic rational $sl(2)$ transfer matrix of the Quantum Inverse Scattering Method \cite{ks79,tf79} with generic inhomogeneities. The conserved operators of \cite{crs97} and the eigenvalues \cite{s00} had in fact appeared in a work of Sklyanin's in 1989 dealing with the problem of separation of variables for Gaudin's spin model \cite{s89}. However this work did not make connection with the $s$-wave pairing Hamiltonian, and it was some time later that the correspondence was realised in full \cite{aff01,o03,vdp02,zlmg02}. It was ultimately shown that the trigonometric analogue is related to the pairing Hamiltonian with $p+ip$-wave pairing symmetry \cite{dilsz10,ilsz09,rdo10,s09}. 

The quasi-classical limit of the Boundary Quantum Inverse Scattering Method was studied by Sklyanin in \cite{s87}, prior to his more well-known publication \cite{s88}. Adopting this approach, several authors have implemented constructions to produce generalised versions of Richardson--Gaudin systems \cite{cmn13,dahog02,h95,s07,s10}. In-depth analyses however, including implications for formulating new pairing Hamiltonians, appear to be have not been widely undertaken. Our study below aims to fill this gap, motivated by a wish to understand the interpretation of the ``boundaries'' in the Richardson--Gaudin context. The broad conclusion from our calculations is that the boundary construction for the spin-$1/2$ case, with the use of diagonal solutions of the reflection equations, does not extend the class of conserved operators beyond results obtained from the twisted-periodic construction. All results for the Bethe Ansatz Equations, the conserved operators, and their eigenvalues can be mapped back, through appropriate changes of variables (and also rescalings and basis transformations in the case of the conserved operators) to analogous quantities obtained from the twisted-periodic formulation. Nonetheless, some counter-intuitive features are uncovered. There is a well-known result of Belavin and Drinfel'd providing a classification of solutions of the quasi-classical Yang--Baxter Equation associated with Lie algebras, in instances where the regularity property holds, into elliptic, trigonometric, and rational cases \cite{bd82}. Our study shows that implementation of the Boundary Quantum Inverse Scattering Method for the Richardson--Gaudin system yields conserved operators whereby the identification of trigonometric and rational parametrisations are interchangeable. We prove that for the Boundary Quantum Inverse Scattering Method formulation in the quasi-classical limit, the rational limit of the trigonometric system is equivalent to the original trigonometric system. Moreover, we prove that the twisted-periodic and boundary constructions are equivalent in the trigonometric case, but not in the rational limit. Some aspects of these equivalences have been previously identified in \cite{dahog02}. Here our aim is to detail a more comprehensive account. 

In Section 2 we begin by introducing a generalised version of Sklyanin's construction using the trigonometric six-vertex solution of the Yang--Baxter Equation, which extends the approach of Karowski and Zapletal \cite{kz94} to include inhomogeneities in the transfer matrix. The algebraic Bethe Ansatz is applied to determine the transfer matrix eigenvalues and associated Bethe Ansatz Equations. This formulation is dependent on a parameter $\rho$ such that Sklyanin's construction is obtained by setting $\rho=0$. In the limit $\rho\rightarrow\infty$ the twisted-periodic transfer matrix is recovered. We refer to this as the {\it attenuated limit}, since it has the effect of collapsing the double-row transfer matrix to the single-row transfer matrix. We also discuss the rational limit, and illustrate the general framework for the well-known case of the Heisenberg $XXZ$ and $XXX$ models.  In Section 3 we turn our attention to a detailed analysis of the quasi-classical limit of this construction. We initially study the Bethe Ansatz Equations in this limit, and establish that several equivalences emerge through appropriately chosen changes of variable. We then show that these same equivalences extend to the conserved operators of the system by identifying appropriate rescalings and basis transformations. Concluding remarks are provided in Section 4. For completeness, we confirm in the Appendix that the equivalences hold at the level of eigenvalue expressions for the conserved operators.   


\section{Boundary Quantum Inverse Scattering Method}\label{sec:BQISM}

In this section we discuss a generalisation of Sklyanin's \textit{Boundary Quantum Inverse Scattering Method} (BQISM) \cite{s88}. A key element is the \textit{$R$-matrix}, which is an invertible operator $R(u)\in\text{End}(V\o V)$ (in this paper $V=\c^2$) depending on the spectral parameter $u\in\c$ and satisfying the \textit{Yang-Baxter Equation} (YBE) \cite{b72,y67}
\begin{equation}\label{YBE}
R_{12}(u-v)R_{13}(u)R_{23}(v)=R_{23}(v)R_{13}(u)R_{12}(u-v).
\end{equation}
It is an equation in $\text{End}(V\o V\o V)$, with the subscripts indicating the spaces in which the corresponding $R$-matrix acts non-trivially.

In this paper we will work with the trigonometric\footnote{While it is conventional to refer to the $R$-matrix as trigonometric, for convenience we adopt the hyperbolic parametrisation.} $R$-matrix associated with the $XXZ$ model \cite{b72}
\begin{equation}\label{R-matrix}
R(u)=\frac{1}{\sinh(u+\eta)}\begin{pmatrix}
                    \sinh(u+\eta) & 0 & 0 & 0\\
                    0 & \sinh u & \sinh\eta & 0\\
                    0 & \sinh\eta & \sinh u & 0\\
                    0 & 0 & 0 & \sinh(u+\eta)\\
                    \end{pmatrix},
\end{equation}
where $\eta\in\c$ is the quasi-classical parameter. Note that (\ref{R-matrix}) satisfies the regularity property, i.e., $R(0)=P$, where $P$ is the permutation operator. 
Also, it is symmetric, i.e., $R_{12}(u)=R_{21}(u)$ and satisfies the unitarity property: $R_{12}(u)R_{12}(-u)=I\o I$.


Noting that $R_{21}^{t_1}(u)$ is invertible, we introduce an additional operator
\begin{equation*}
\mathcal{R}_{12}(u)=\left(\left(R_{21}^{t_1}(u)\right)^{-1}\right)^{t_1}\in\text{End}(V\o V),
\end{equation*}
where $t_1$ denotes the partial transpose over the first space in the tensor product. One can check that for the trigonometric $R$-matrix $\mathcal{R}(u)\propto R(-u-2\eta)$. 
\begin{remark}
By construction, 
\begin{equation*}
\mathcal{R}_{12}^{t_1}(u)R_{21}^{t_1}(u)=\mathcal{R}_{21}^{t_2}(u)R_{12}^{t_2}(u)=I\o I.
\end{equation*}
\end{remark}

In the BQISM framework we require that in addition to the YBE (\ref{YBE}) the $R$-matrix satisfies two \textit{reflection equations} in $\text{End}(V\o V)$ \cite{c84}
\begin{equation}\label{reflection}
\begin{aligned}
R_{12}(u-v)K_1^-(u)R_{21}(u+v)K_2^-(v)&=K_2^-(v)R_{12}(u+v)K_1^-(u)R_{21}(u-v),\\
R_{12}(v-u)K_1^+(u)\mathcal{R}_{21}(u+v)K_2^+(v)&=K_2^+(v)\mathcal{R}_{12}(u+v)K_1^+(u)R_{21}(v-u)
\end{aligned}
\end{equation}
for some operators $K^\pm\in\text{End}(V)$, referred to as the \textit{reflection matrices} or the \textit{$K$-matrices}.
One can check that the following reflection matrices satisfy equations (\ref{reflection}) together with the trigonometric $R$-matrix (\ref{R-matrix}):
\begin{equation}\label{K-matrices_0}
\begin{aligned}
K^-(u)&=\begin{pmatrix}
        \sinh(\xi^- +u) & 0 \\
        0 & \sinh(\xi^- -u)
        \end{pmatrix}, \\
K^+(u)&=\begin{pmatrix}
        \sinh(\xi^+ +u+\eta) & 0 \\
        0 & \sinh(\xi^+ -u-\eta)
        \end{pmatrix}.                     
\end{aligned}
\end{equation}
Introduce the \textit{double row monodromy matrix} acting in $V_a\o V^{\o\l}$, where $V_a$ is called the \textit{auxiliary space} (in our case a copy of $\c^2$) and $V^{\o\l}$ is the \textit{quantum space},
\begin{equation}\label{T-matrix_0}
T_a(u)=R_{a\l}(u-\eps_{\l})... R_{a1}(u-\eps_1)K_a^-(u+\rho/2)R_{a1}(u+\eps_1+\rho)... R_{a\mathcal{L}}(u+\eps_{\l}+\rho),
\end{equation}
where $\rho,\eps_j\in\c$ are complex parameters. The parameters $\eps_j$ are known as \textit{inhomogeneities}. These are typically set to be zero in the construction of 
one-dimensional quantum lattice models, but are retained as generic parameters in Richardson--Gaudin systems.

Using (\ref{YBE}) one can check that the monodromy matrix $T(u)$ given by (\ref{T-matrix_0}) satisfies the following reflection type equation in $V_a\o V_b\o V^{\o\mathcal{L}}$:
\begin{equation}\label{RTRT}
R_{ab}(u-v)T_a(u)R_{ba}(u+v+\rho)T_b(v)=T_b(v)R_{ab}(u+v+\rho)T_a(u)R_{ba}(u-v).
\end{equation}

\begin{remark}
We are implementing a modification of Sklyanin's formulation, following Karowski and Zapletal \cite{kz94}. This consists of introducing an additional parameter $\rho$, which provides a shift in the parameters: 
$u\mapsto u+{\rho}/{2},\ \eps_l\mapsto \eps_l+{\rho}/{2}$. It will allow us to interpolate between the boundary and the twisted-periodic cases. The limit $\rho\rightarrow 0$ reduces to the boundary formulation, while the limit $\rho\rightarrow\infty$, as we will see later, yields the twisted-periodic construction.
\end{remark}
The next step is to introduce the \textit{transfer matrix}
\begin{equation}\label{t-matrix_0}
t(u)={\rm tr}_a\left(K_a^+(u+\rho/2)T_{a}(u)\right).
\end{equation}
Using (\ref{RTRT}) one can prove that the transfer matrices given by (\ref{t-matrix_0}) commute for any two values of the spectral parameter:
\begin{equation*}
[t(u),t(v)]=0 \text{\ \ for all\ \ }u,v\in\c.
\end{equation*}
This is a fundamental property of the transfer matrix that allows it to be used it as a generating function for the conserved operators.

For future calculations it is convienient to introduce another shift $u\mapsto u-{\eta}/{2}$ in the spectral parameter and to redefine all functions taking this into account. It is also convenient to introduce the \textit{Lax operator} obtained as a scaling of the (shifted) $R$-matrix:
\begin{equation}\label{L-operator}
\begin{aligned}
\check{L}(u)=&\,\frac{\sinh(u+\eta/2)}{\sinh u}R(u-\eta/2)=\\
=&\,\frac{1}{\sinh u}\begin{pmatrix}
                 \sinh(u+\eta/2) & 0 & 0 & 0\\
                 0 & \sinh (u-\eta/2) & \sinh\eta & 0\\
                 0 & \sinh\eta & \sinh (u-\eta/2) & 0\\
                 0 & 0 & 0 & \sinh(u+\eta/2)\\
                 \end{pmatrix}.
\end{aligned}                 
\end{equation}
It satisfies the YBE
\begin{equation*}
R_{12}(u-v)\check{L}_{13}(u)\check{L}_{23}(v)=\check{L}_{23}(v)\check{L}_{13}(u)R_{12}(u-v).
\end{equation*}
Also, we need to rescale the $K$-matrices (\ref{K-matrices_0}):
\begin{equation}\label{K-matrices}
\begin{aligned}
\check{K}^-(u)&=\frac{1}{\sinh u}K^-(u-\eta/2)=\frac{1}{\sinh u}
                                             \begin{pmatrix}
                                             \sinh(\xi^- +u-\eta/2) & 0 \\
                                             0 & \sinh(\xi^- -u+\eta/2)
                                             \end{pmatrix}, \\
\check{K}^+(u)&=\frac{1}{\sinh u}K^+(u-\eta/2)=\frac{1}{\sinh u}
                                             \begin{pmatrix}
                                             \sinh(\xi^+ +u+\eta/2) & 0 \\
                                             0 & \sinh(\xi^+ -u-\eta/2)
                                             \end{pmatrix}.                 
\end{aligned}
\end{equation}
The monodromy matrix is now
\begin{equation}\label{T-matrix}
\check{T}_a(u)=\check{L}_{a\l}(u-\eps_{\l})... \check{L}_{a1}(u-\eps_1)\check{K}_a^-(u+\rho/2)\check{L}_{a1}(u+\eps_1+\rho)... \check{L}_{a\l}(u+\eps_{\l}+\rho),
\end{equation}
and the transfer matrix is, correspondingly,
\begin{equation}\label{t-matrix}
\check{t}(u)={\rm tr}_a\left(\check{K}_a^+(u+\rho/2)\check{T}_{a}(u)\right).
\end{equation}
One can write the monodromy matrix (\ref{T-matrix}) as an operator valued $2\times 2$-matrix in the auxiliary space:
\begin{equation*}
\check{T}_a(u)=\begin{pmatrix}
             A(u) & B(u) \\
             C(u) & D(u)
             \end{pmatrix}.
\end{equation*}
It is convenient to work with $\tilde{A}(u)=\sinh(2u+\rho) A(u)-\sinh\eta D(u)$ instead of $A(u)$. Then, using (\ref{RTRT}), one can show that the following commutation relations hold:
\begin{equation}\label{ACDC}
\begin{aligned}
D(u)C(v)=&\,\frac{\sinh(u-v-\eta)\sinh(u+v+\rho-\eta)}{\sinh(u-v)\sinh(u+v+\rho)}C(v)D(u)
+\frac{\sinh\eta\sinh(2v+\rho-\eta)}{\sinh(u-v)\sinh(2v+\rho)}C(u)D(v)\,-\\
-&\,\frac{\sinh\eta}{\sinh(2v+\rho)\sinh(u+v+\rho)}C(u)\tilde{A}(v),\\
\tilde{A}(u)C(v)=&\,\frac{\sinh(u-v+\eta)\sinh(u+v+\rho+\eta)}{\sinh(u-v)\sinh(u+v+\rho)}C(v)\tilde{A}(u)
-\frac{\sinh\eta\sinh(2u+\rho+\eta)}{\sinh(u-v)\sinh(2v+\rho)}C(u)\tilde{A}(v)\,+\\
+&\,\frac{\sinh\eta\sinh(2v+\rho-\eta)\sinh(2u+\rho+\eta)}{\sinh(u+v+\rho)\sinh(2v+\rho)}C(u)D(v).
\end{aligned}
\end{equation}
The transfer matrix (\ref{t-matrix}) can be written in the form
\begin{equation*}
\check{t}(u)=\frac{\sinh(\xi^+ +u+\rho/2+\eta/2)}{\sinh(2u+\rho)\sinh(u+\rho/2)}\tilde{A}(u)+\frac{\sinh(2u+\rho+\eta)\sinh(\xi^+ -u-\rho/2+\eta/2)}{\sinh(2u+\rho)\sinh(u+\rho/2)}D(u).
\end{equation*}
To find it's eigenstates and eigenvalues we follow the \textit{algebraic Bethe Ansatz} as described in \cite{s88}. We start with a reference state $\Omega\in V^{\o\mathcal{L}}$, s.t.
\begin{equation*}
B(u)\Omega=0,\ A(u)\Omega=a(u)\Omega,\ D(u)\Omega=d(u)\Omega,\ C(u)\Omega\neq 0,
\end{equation*}
where $a(u)$ and $d(u)$ are scalar functions, so that $\Omega$ is an eigenstate for $A(u)$ and $D(u)$ simultaneously and, hence, also for $\tilde{A}(u)$:
$\tilde{A}(u)\Omega=\tilde{a}(u)\Omega$, where $\tilde{a}(u)=\sinh(2u+\rho)a(u)-\sinh\eta d(u)$. Thus, it is also an eigenstate of $\check{t}(u)$, which is a linear combination of $\tilde{A}(u)$ and $D(u)$. It is an analogue to a ``lowest weight'' state in the representation theory of $\mathfrak{gl}(2)$.

We next look for other eigenstates in the form
\begin{equation}\label{phi}
\Phi=\Phi(v_1,...,v_N)=C(v_1)...C(v_N)\Omega.
\end{equation}
Using relations (\ref{ACDC}) one can prove that the state $\Phi$ given by (\ref{phi}) is an eigenstate of $\check{t}(u)$ with the eigenvalue
\begin{equation}\label{lambda}
\begin{aligned}
\check{\Lambda}&(u,v_1,...,v_N)=\tilde{a}(u)\frac{\sinh(\xi^+ +u+\rho/2+\eta/2)}{\sinh(2u+\rho)\sinh(u+\rho/2)}
\prod_{j=1}^N\frac{\sinh(u-v_j+\eta)\sinh(u+v_j+\rho+\eta)}{\sinh(u-v_j)\sinh(u+v_j+\rho)}\,+\\
+&\,d(u)\frac{\sinh(2u+\rho+\eta)\sinh(\xi^+ -u-\rho/2+\eta/2)}{\sinh(2u+\rho)\sinh(u+\rho/2)}
\prod_{j=1}^N\frac{\sinh(u-v_j-\eta)\sinh(u+v_j+\rho-\eta)}{\sinh(u-v_j)\sinh(u+v_j+\rho)},
\end{aligned}
\end{equation}
if $\Phi\neq 0$ and the \textit{Bethe Ansatz Equations} (BAE) are satisfied:
\begin{equation}\label{bae}
\frac{\tilde{a}(v_k)}{d(v_k)\sinh(2v_k+\rho-\eta)}\frac{\sinh(\xi^+ +v_k+\rho/2+\eta/2)}{\sinh(\xi^+ -v_k-\rho/2+\eta/2)}=
\prod_{j\neq k}^N\frac{\sinh(v_k-v_j-\eta)\sinh(v_k+v_j+\rho-\eta)}{\sinh(v_k-v_j+\eta)\sinh(v_k+v_j+\rho+\eta)}.
\end{equation}
One can check that $\Omega=\begin{pmatrix}
                           0 \\
                           1 
                           \end{pmatrix}^{\o\l}$ is a reference state. Then 
\begin{equation*}
\check{L}_{al}(u-\eps_l)
        \begin{pmatrix}
         0 \\
         1 
        \end{pmatrix}_l=
    \dfrac{1}{\sinh(u-\eps_l)}
    \begin{pmatrix}
     \sinh(u-\eps_l-\eta/2) & 0 \\
     * & \sinh(u-\eps_l+\eta/2)
    \end{pmatrix}
    \begin{pmatrix}
         0 \\
         1 
    \end{pmatrix}_l,
\end{equation*}

\begin{equation*}
\check{L}_{al}(u+\eps_l+\rho)
        \begin{pmatrix}
         0 \\
         1 
        \end{pmatrix}_l=
    \dfrac{1}{\sinh(u+\eps_l+\rho)}
    \begin{pmatrix}
     \sinh(u+\eps_l+\rho-\eta/2) & 0 \\
     * & \sinh(u+\eps_l+\rho+\eta/2)
    \end{pmatrix}
    \begin{pmatrix}
         0 \\
         1 
    \end{pmatrix}_l
\end{equation*}
where we follow the tradition that $*$ denotes an operator which does need to be known to continue calculations. From here one can derive the formulae for $\tilde{a}(u)$ and $d(u)$:   
\begin{equation}\label{ad}                        
\begin{aligned}
\tilde{a}(u)=&\,\sinh(2u+\rho-\eta)\frac{\sinh(\xi^- +u+\rho/2+\eta/2)}{\sinh(u+\rho/2)}
\prod_{l=1}^{\mathcal{L}}\frac{\sinh(u-\eps_l-\eta/2)\sinh(u+\eps_l+\rho-\eta/2)}{\sinh(u-\eps_l)\sinh(u+\eps_l+\rho)},\\
d(u)=&\,\frac{\sinh(\xi^- -u-\rho/2+\eta/2)}{\sinh(u+\rho/2)}
\prod_{l=1}^{\mathcal{L}}\frac{\sinh(u-\eps_l+\eta/2)\sinh(u+\eps_l+\rho+\eta/2)}{\sinh(u-\eps_l)\sinh(u+\eps_l+\rho)}.
\end{aligned}
\end{equation}

In the following, we look to take various limits of quantities such as the operators $\check{K}^\pm(u)$ and $\check{L}(u)$, the transfer matrix, its eigenvalues and the BAE. For readability we have chosen not to introduce new notation for each limiting object, but will ensure that it is clear which expression is being affected. 


\subsection{Attenuated limit}

Setting $\rho=0$ above, the construction reduces to the regular form of the BQISM with inhomogeneities. In this section we show that the limit $\rho\rightarrow\infty$ reduces to the twisted-periodic QISM formulation, where the twist is sector dependent. We refer to this limit as the \textit{attenuated limit}, since the double row transfer matrix reduces to a single row transfer matrix as  $\rho\rightarrow\infty$. This approach was used in \cite{kz94} to construct twisted-periodic one-dimensional quantum lattice models in a manner which preserved certain Hopf-algebraic symmetries.  


Substituting the expression (\ref{T-matrix}) for $\check{T}_a(u)$, we may explicitly write the transfer matrix (\ref{t-matrix}) as 
\begin{equation}\label{t-matrix-2}
\check{t}(u)={\rm tr}_a\Big(\check{K}^+_a(u+\rho/2)\check{L}_{a\l}(u-\eps_{\l}) ... \check{L}_{a1}(u-\eps_1)
\check{K}_a^-(u+\rho/2)\check{L}_{a1}(u+\eps_1+\rho)... \check{L}_{a\l}(u+\eps_{\l}+\rho)\Big).
\end{equation}
We have
\begin{equation*}
\check{L}(u)\ \xrightarrow{u\rightarrow\infty}\        
                 M=\begin{pmatrix}
                   q^{1/2} & 0 & 0 & 0\\
                   0 & q^{-1/2} & 0 & 0\\
                   0 & 0 & q^{-1/2} & 0\\
                   0 & 0 & 0 & q^{1/2}\\
                   \end{pmatrix},
\end{equation*}
where $q=\exp\eta$.

Consider a matrix $\hat{N}_j=\begin{pmatrix}
                           1 & 0 \\
                           0 & 0
                           \end{pmatrix}_j$ acting on the $j$th $V$ space from the tensor product $V^{\o\l}$. We then have
\begin{equation*}
\begin{pmatrix}
q^{1/2} & 0 \\
0 & q^{-1/2}
\end{pmatrix}_j=q^{\hat{N}_j-1/2},\ \ \ 
\begin{pmatrix}
q^{-1/2} & 0 \\
0 & q^{1/2}
\end{pmatrix}_j=q^{1/2-\hat{N}_j}.
\end{equation*}   
Thus,
\begin{equation*}
\check{L}_{aj}(u)\ \xrightarrow{u\rightarrow\infty}\ 
M_j=\begin{pmatrix}
    q^{\hat{N}_j-1/2} & 0 \\
    0 & q^{1/2-\hat{N}_j}
    \end{pmatrix},
\end{equation*}  
and 
\begin{equation*}
\begin{aligned}
&\check{L}_{a1}(u+\eps_1+\rho)... \check{L}_{a\l}(u+\eps_{\l}+\rho)\ \xrightarrow{\rho\rightarrow\infty}\ M_1 M_2...M_\l=\\
=&\begin{pmatrix}
  q^{\hat{N}_1-1/2} & 0 \\
  0 & q^{1/2-\hat{N}_1}
  \end{pmatrix}...\begin{pmatrix}
                  q^{\hat{N}_\l-1/2} & 0 \\
                  0 & q^{1/2-\hat{N}_\l}
                  \end{pmatrix}=\begin{pmatrix}
                                q^{\hat{N}-\l/2} & 0 \\
                                0 & q^{\l/2-\hat{N}}
                                \end{pmatrix},
\end{aligned}
\end{equation*}  
where 
$\displaystyle \hat{N}=\sum_{l=1}^\l \hat{N}_l$. A transfer matrix eigenstate $\Phi$ is also an eigenstate of the operator $\hat{N}$ with eigenvalue  equal to $N$, the number of $C$-operators applied to the reference state in order to obtain $\Phi=C(v_1)...C(v_N)\Omega$. In this manner it is seen that the transfer matrix has a block diagonal structure whereby $\hat{N}$ takes a constant value on each block.

Furthermore,
\begin{equation*}
\begin{aligned}
\check{K}^-(u)&=\frac{1}{\sinh u}\begin{pmatrix}
                               \sinh(\xi^- +u-\eta/2) & 0 \\
                               0 & \sinh(\xi^- -u+\eta/2)
                               \end{pmatrix}\ \xrightarrow{u\rightarrow\infty}\ 
                               \begin{pmatrix}
                               e^{\xi^- -\eta/2} & 0 \\
                               0 & -e^{-\xi^- -\eta/2}
                               \end{pmatrix},\\
\check{K}^+(u)&=\frac{1}{\sinh u}\begin{pmatrix}
                               \sinh(\xi^+ +u+\eta/2) & 0 \\
                               0 & \sinh(\xi^+ -u-\eta/2)
                               \end{pmatrix}\ \xrightarrow{u\rightarrow\infty}\    
                               \begin{pmatrix}
                               e^{\xi^+ +\eta/2} & 0 \\
                               0 & -e^{-\xi^+ +\eta/2}
                               \end{pmatrix}.                           
\end{aligned}
\end{equation*}   
Denote
\begin{equation*}
\check{L}_{a\l}(u-\eps_{\l})... \check{L}_{a1}(u-\eps_1)=\begin{pmatrix}
                                                         A_1 & B_1 \\
                                                         C_1 & D_1
                                                         \end{pmatrix}.
\end{equation*} 
We then have 
\begin{equation*}
\begin{aligned}
\check{t}(u)\ \xrightarrow{\rho\rightarrow\infty}\ &{\rm tr}_a\left(
                              \begin{pmatrix}
                               e^{\xi^+ +\eta/2} & 0 \\
                               0 & -e^{-\xi^+ +\eta/2}
                               \end{pmatrix}_a\begin{pmatrix}
                                            A_1 & B_1 \\
                                            C_1 & D_1
                                            \end{pmatrix}\begin{pmatrix}
                                                         e^{\xi^- -\eta/2} & 0 \\
                                                         0 & -e^{-\xi^- -\eta/2}
                                                         \end{pmatrix}_a\begin{pmatrix}
                                                                      q^{\hat{N}-\l/2} & 0 \\
                                                                      0 & q^{\l/2-\hat{N}}
                                                                      \end{pmatrix}\right)=\\
=&\exp(\xi^+ +\xi^-)A_1 \exp\eta(\hat{N}-\l/2)+\exp(-\xi^+ -\xi^-)D_1\exp\eta(\l/2-\hat{N}).
\end{aligned}
\end{equation*}
Since $\hat{N}$ is a conserved operator, it commutes with both $A_1$ and $D_1$. Thus,
\begin{equation}\label{t-matrix_limit}
\check{t}(u)\ \xrightarrow{\rho\rightarrow\infty}\ \exp(\xi^+ +\xi^- +\eta N-\eta\l/2)A_1+\exp(-\xi^+ -\xi^-+\eta\l/2-\eta N)D_1.
\end{equation}

\begin{remark}
The twisted-periodic transfer matrix has the form \cite{s89}
\begin{equation}\label{t-matrix_periodic}
t(u)={\rm tr}_a\left(\begin{pmatrix}
               e^{-\eta\gamma} & 0 \\
               0 & e^{\eta\gamma}
               \end{pmatrix}_a\check{L}_{a\l}(u-\eps_{\l})... \check{L}_{a1}(u-\eps_1)\right)=\exp(-\eta\gamma)A_1+\exp(\eta\gamma)D_1.
\end{equation}
Thus, to obtain the twisted-periodic transfer matrix (\ref{t-matrix_periodic}) from the attenuated limit (\ref{t-matrix_limit}) of the boundary transfer matrix (\ref{t-matrix}), we need to impose that $\gamma$ depends on $N$:
\begin{equation}\label{gamma}
\gamma=\l/2-N-\eta^{-1}(\xi^+ +\xi^-).
\end{equation}
\end{remark}



From (\ref{ad}) we can compute that
\begin{equation*}
\frac{\tilde{a}(v_k)}{d(v_k)\sinh(2v_k+\rho-\eta)}=\frac{\sinh(\xi^- +v_k+\rho/2+\eta/2)}{\sinh(\xi^- -v_k-\rho/2+\eta/2)}
\prod_{l=1}^{\l}\frac{\sinh(v_k-\eps_l-\eta/2)\sinh(v_k+\eps_l+\rho-\eta/2)}{\sinh(v_k-\eps_l+\eta/2)\sinh(v_k+\eps_l+\rho+\eta/2)}.
\end{equation*}
In the limit as $\rho\rightarrow\infty$:
\begin{equation*}
\begin{aligned}
&\frac{\sinh(\xi^- +v_k+\rho/2+\eta/2)}{\sinh(\xi^- -v_k-\rho/2+\eta/2)}\ \xrightarrow{\rho\rightarrow\infty}\ -\exp(2\xi^- +\eta),\\
&\frac{\sinh(\xi^+ +v_k+\rho/2+\eta/2)}{\sinh(\xi^+ -v_k-\rho/2+\eta/2)}\ \xrightarrow{\rho\rightarrow\infty}\ -\exp(2\xi^+ +\eta),\\
&\frac{\sinh(v_k+\eps_l+\rho-\eta/2)}{\sinh(v_k+\eps_l+\rho+\eta/2)}\ \xrightarrow{\rho\rightarrow\infty}\ \exp(-\eta),\\
&\frac{\sinh(v_k+v_j+\rho-\eta)}{\sinh(v_k+v_j+\rho+\eta)}\ \xrightarrow{\rho\rightarrow\infty}\ \exp(-2\eta).
\end{aligned}
\end{equation*}
Thus, the BAE (\ref{bae}) in this limit reduce to
\begin{equation}\label{bae_limit}
\exp(2(\xi^+ +\xi^-)-\eta\l+2\eta N)\prod_{l=1}^{\l}\frac{\sinh(v_k-\eps_l-\eta/2)}{\sinh(v_k-\eps_l+\eta/2)}=\prod_{j\neq k}^N\frac{\sinh(v_k-v_j-\eta)}{\sinh(v_k-v_j+\eta)}.
\end{equation}
In a similar manner we obtain the limit of (\ref{lambda}) as
\begin{equation}\label{lambda_limit}
\begin{aligned}
\check{\Lambda}(u)&\xrightarrow{\rho\rightarrow\infty}\exp(\xi^+ +\xi^- -\eta\l/2+\eta N)\prod_{l=1}^{\mathcal{L}}\frac{\sinh(u-\eps_l-\eta/2)}{\sinh(u-\eps_l)}\prod_{j=1}^N\frac{\sinh(u-v_j+\eta)}{\sinh(u-v_j)}+\\
&\quad +\exp(-\xi^+ -\xi^- +\eta\l/2-\eta N)\prod_{l=1}^{\mathcal{L}}\frac{\sinh(u-\eps_l+\eta/2)}{\sinh(u-\eps_l)}\prod_{j=1}^N\frac{\sinh(u-v_j-\eta)}{\sinh(u-v_j)}.
\end{aligned}
\end{equation}
\begin{remark}
We recognise that (\ref{bae_limit}) subject to (\ref{gamma}) are the BAE for (\ref{t-matrix_periodic}), as required; e.g. see \cite{dilsz10,vdp02}. We also recognise that (\ref{lambda_limit}) subject to (\ref{gamma}) are the eigenvalues of (\ref{t-matrix_periodic}).
\end{remark}

\subsection{Rational limit}\label{sec:rational_limit}

In this section we show that there is a relationship between the rational twisted-periodic system and the
rational boundary system that is similar to the trigonometric case that we have just
discussed in the previous section. By introducing a parameter $\nu$ (the so-called {\em rational
parameter}) as a scaling factor in the argument of the hyperbolic functions, and using 
$\displaystyle{\lim_{\nu\to 0}\frac{\sinh(\nu x)}{\nu}=x}$, 
one can obtain the {\em rational limit} of the
relevant operators $\check{L}(u)$ of equation (\ref{L-operator}) and the $\check{K}^\pm(u)$ of equations
(\ref{K-matrices}) as follows:
\begin{equation}\label{RatL}
\check{L}(u)\rightarrow\frac{1}{u}
                 \begin{pmatrix}
                 u+\eta/2 & 0 & 0 & 0\\
                 0 & u-\eta/2 & \eta & 0\\
                 0 & \eta & u-\eta/2 & 0\\
                 0 & 0 & 0 & u+\eta/2\\
                 \end{pmatrix},
\end{equation}

\begin{align}
\check{K}^-(u)&\rightarrow\frac{1}{u}
              \begin{pmatrix}
              \xi^- +u-\eta/2 & 0 \\
              0 & \xi^- -u+\eta/2
              \end{pmatrix}, \label{RatKm}\\
\check{K}^+(u)&\rightarrow\frac{1}{u}
              \begin{pmatrix}
              \xi^+ +u+\eta/2 & 0 \\
              0 & \xi^+ -u-\eta/2
              \end{pmatrix}. \label{RatKp}    
\end{align}
We observe that in this same limit, the BAE (\ref{bae}) become
\begin{equation}\label{RatBAE}
\begin{aligned}
&\frac{(\xi^-+v_k+\rho/2+\eta/2)(\xi^++v_k+\rho/2+\eta/2)}{(\xi^--v_k-\rho/2+\eta/2)(\xi^+-v_k-\rho/2+\eta/2)}
\prod_{l=1}^\l\frac{(v_k-\eps_l-\eta/2)(v_k+\eps_l+\rho-\eta/2)}{(v_k-\eps_l+\eta/2)(v_k+\eps_l+\rho+\eta/2)} =\\
&\qquad = \prod_{j\neq k}^N\frac{(v_k-v_j-\eta)(v_k+v_j+\rho-\eta)}{(v_k-v_j+\eta)(v_k+v_j+\rho+\eta)}, 
\end{aligned}
\end{equation}
and the expression for the eigenvalues given in (\ref{lambda}) reduces to 
\begin{equation}\label{RatEval}
\begin{aligned}
\check{\Lambda}(u,v_1,...,v_N)
&\rightarrow 
\frac{(u+\rho/2-\eta/2)(\xi^-+u+\rho/2+\eta/2)(\xi^++u+\rho/2+\eta/2)}{(u+\rho/2)^3} \times \\
&\quad \times \prod_{l=1}^\l\frac{(u-\eps_l-\eta/2)(u+\eps_l+\rho-\eta/2)}{(u-\eps_l)(u+\eps_l+\rho)}
\prod_{j=1}^N\frac{(u-v_j+\eta)(u+v_j+\rho+\eta)}{(u-v_j)(u+v_j+\rho)}+ \\ 
&\quad + \frac{(u+\rho/2+\eta/2)(\xi^--u-\rho/2+\eta/2)(\xi^+-u-\rho/2+\eta/2)}{(u+\rho/2)^3}\times \\
&\quad \times \prod_{l=1}^\l\frac{(u-\eps_l+\eta/2)(u+\eps_l+\rho+\eta/2)}{(u-\eps_l)(u+\eps_l+\rho)}
\prod_{j=1}^N\frac{(u-v_j-\eta)(u+v_j+\rho-\eta)}{(u-v_j)(u+v_j+\rho)}.
\end{aligned}
\end{equation}

The transfer matrix (\ref{t-matrix}) in the rational limit, particularly in
the form (\ref{t-matrix-2}), is readily obtained by employing the expressions
(\ref{RatL}), (\ref{RatKm}) and (\ref{RatKp}) above. To then determine the attenuated limit of this
rational transfer matrix, we first observe that from (\ref{RatL}) above, $\check{L}(u)\rightarrow I$
as $u\rightarrow\infty$. This implies that the terms $\check{L}_{aj}(u+\eps_j+\rho)$ occuring
to the right of $\check{K}^-_a(u+\rho/2)$ in (\ref{t-matrix-2}) all simplify to the identity as
$\rho\rightarrow\infty$. 
Without loss of generality, we moreover suppose that $\xi^-$ does not depend on $\rho$, in which case taking the
attenuated limit of (\ref{RatKm}) gives
\begin{equation*}
\check{K}^-(u+\rho/2)\ \xrightarrow{\rho\rightarrow\infty}\ \begin{pmatrix}
                                                            1 & 0 \\
                                                            0 & -1
                                                            \end{pmatrix}.
\end{equation*} 
Furthermore, we set $\xi^+=\zeta\rho$, where $\zeta\in\c$, from which we obtain the attenuated limit
of equation (\ref{RatKp}) above:
\begin{equation*}
\check{K}^+(u+\rho/2)\ \xrightarrow{\rho\rightarrow\infty}\ \begin{pmatrix}
                                                            2\zeta+1 & 0 \\
                                                            0 & 2\zeta-1
                                                            \end{pmatrix}.
\end{equation*}
Thus, we have the attenuated limit of the rational transfer matrix in the form (\ref{t-matrix-2}) being given by 
\begin{equation*}
\check{t}(u)\ \xrightarrow{\rho\rightarrow\infty}\ {\rm tr}_a\left(\begin{pmatrix}
                                                             2\zeta+1 & 0 \\
                                                             0 & 2\zeta-1
                                                             \end{pmatrix}_a\check{L}_{a\l}(u-\eps_{\l})... \check{L}_{a1}(u-\eps_1)\begin{pmatrix}
                                                                                                                                  1 & 0 \\
                                                                                                                                  0 & -1
                                                                                                                                  \end{pmatrix}_a\right)=
(1+2\zeta)A_1+(1-2\zeta)D_1,
\end{equation*}
where the operators $\check{L}_{aj}(u-\eps_j)$ and, correspondingly, the operators $A_1$ and $D_1$ are in the rational limit.

Finally, imposing the condition that $\zeta\neq \pm1/2$ to avoid any technical issues of divergence, 
for convenience we rescale
\begin{equation*}
\check{K}^+(u+\rho/2)\rightarrow\dfrac{1}{\sqrt{1-4\zeta^2}}\check{K}^+(u+\rho/2)
\end{equation*}
to match this limiting expression for $\check{t}(u)$ with that of the twisted-periodic case given in equation (\ref{t-matrix_periodic}) above. This is achieved by setting
\begin{equation}\label{gamma-rat}
e^{-\eta\gamma}=\frac{1+2\zeta}{\sqrt{1-4\zeta^2}},\ \ e^{\eta\gamma}=\frac{1-2\zeta}{\sqrt{1-4\zeta^2}}.
\end{equation}
In the attenuated limit (i.e. $\rho\to\infty$), the rational BAE (\ref{RatBAE}) become 
\begin{equation}\label{AttRatBAE}
\frac{1+2\zeta}{1-2\zeta}\prod_{l=1}^{\l}\frac{v_k-\eps_l-\eta/2}{v_k-\eps_l+\eta/2}=\prod_{j\neq k}^N\frac{v_k-v_j-\eta}{v_k-v_j+\eta}.
\end{equation}
It is evident that by setting 
\begin{equation}\label{gamma-att-rat}
e^{-2\eta\gamma}=\dfrac{1+2\zeta}{1-2\zeta},
\end{equation} 
we may identify (\ref{AttRatBAE}) with the rational limit of (\ref{bae_limit}). It is also worth pointing out that (\ref{gamma-att-rat}) is consistent with (\ref{gamma-rat}).
	
Finally, the expression for the eigenvalues (\ref{RatEval}) in the attenuated limit is  
\begin{equation}
\check{\Lambda}(u,v_1,...,v_N)\rightarrow\frac{1+2\zeta}{\sqrt{1-4\zeta^2}}\prod_{l=1}^{\mathcal{L}}\frac{u-\eps_l-\eta/2}{u-\eps_l}\prod_{j=1}^N\frac{u-v_j+\eta}{u-v_j}+
\frac{1-2\zeta}{\sqrt{1-4\zeta^2}}\prod_{l=1}^{\mathcal{L}}\frac{u-\eps_l+\eta/2}{u-\eps_l}\prod_{j=1}^N\frac{u-v_j-\eta}{u-v_j}.
\label{AttRatEval}
\end{equation}
By once again applying (\ref{gamma-rat}), we may identify the expression (\ref{AttRatEval}) with the
rational limit of (\ref{lambda_limit}). In other words, we have shown that the rational and
attenuated limits {\em commute}, subject to appropriate scaling of relevant quantities.

A convenient way to summarise our discussions so far in Section \ref{sec:BQISM} is to provide
a diagram highlighting the connections we have made between the various trigonometric, hereafter denoted \textbf{Trig.}, and rational, hereafter denoted \textbf{Rat.},
limits. We will also use the notations \textbf{BQISM} to denote the general construction, and \textbf{QISM} for the attentuated limit.     
Note below that \textbf{Trig.\ BQISM$^\prime$} and \textbf{Rat.\ BQISM$^\prime$} are merely the respective
\textbf{Trig.\ BQISM} and \textbf{Rat.\ BQISM} with $\rho$ included explicitly in all expressions.
We do not consider these to be fundamentally different systems (consider variable change $\#1$ in
the diagram, denoted simply by $\#1$, which
is just $v_k\mapsto v_k+{\rho}/{2},\ \eps_l\mapsto \eps_l+{\rho}/{2}$), but make the
distinction as a convenience to highlight our utilisation of the methods of Karowski and Zapletal
\cite{kz94} via the attenuated limit.
$$
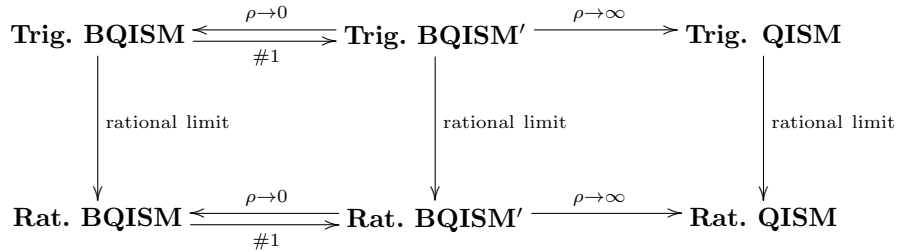
\xymatrix{
{\txt{\textbf{Trig.\ BQISM}}}  \ar[dd]^{\rm rational\ limit}  \ar@<-1ex>[rr]_{\#1}  && \ar[ll]_{\rho\rightarrow 0} 
{\txt{\textbf{Trig.\ BQISM$^\prime$}}} \ar[rr]^{\rho\rightarrow\infty} \ar[dd]^{\rm rational\ limit} &&
  {\txt{\textbf{Trig. QISM}}}   \ar[dd]^{\rm rational\ limit} \\
  && \\ 
{\txt{\textbf{Rat.\ BQISM}}} \ar@<-1ex>[rr]_{\#1} && \ar[ll]_{\rho\rightarrow 0} {\txt{\textbf{Rat.\ BQISM$^\prime$}}}  \ar[rr]^{\rho\rightarrow\infty} &&
{\txt{\textbf{Rat.\ QISM}}}               
}
$$


\subsection{Heisenberg model} \label{sec:hm}

In this section we show how the Heisenberg model can be obtained as a special case from the general construction outlined so far. Here we will omit the shift $u\mapsto u-\eta/2$ and the scalings described in equations (\ref{L-operator}) - (\ref{t-matrix}), in order to obtain the standard form of the Heisenberg model.

Consider the transfer matrix (\ref{t-matrix_0}) with $\eps_j=0$: 
\begin{equation}\label{heisenberg_rho}
t(u)={\rm tr}_a\left(K_a^+(u+\rho/2)R_{a\l}(u)... R_{a1}(u)K_a^-(u+\rho/2)R_{a1}(u+\rho)...R_{a\l}(u+\rho)\right).
\end{equation}
If we take $\rho\rightarrow0$ we obtain the open chain Heisenberg model transfer matrix:
\begin{equation}\label{heisenberg_open}
t(u)\rightarrow {\rm tr}_a\left(K_a^+(u)R_{a\l}(u)... R_{a1}(u)K_a^-(u)R_{a1}(u)...R_{a\l}(u)\right).
\end{equation}
The Hamiltonian is constructed from $t(u)$ given by (\ref{heisenberg_open}) as follows:
\begin{equation}\label{openXXZ}
H=t^{-1}(0)t'(0)=\sum_{j=1}^{\l-1}H_{j(j+1)}+\frac{1}{2} \left(K_1^-(0)\right)^{-1} \left(K_1^-\right)'(0) 
+\frac{{\rm tr}_a\left(K_a^+(0)H_{a\l}\right)}{{\rm tr}_a\left(K_a^+(0)\right)},
\end{equation}
where $H_{j(j+1)}=P_{j(j+1)}R'_{j(j+1)}(0),\ H_{a\l}=R_{a\l}'(0)P_{a\l}$, and $t'(0)$,
$R'_{j(j+1)}(0)$ and $\left(K_1^-\right)'(0)$ are derivatives of the corresponding operators at
$u=0$. The explicit form of the Hamiltonian (\ref{openXXZ}) in terms of Pauli matrices may be found in \cite{s88}.

Now if we consider $\rho\rightarrow\infty$, the transfer matrix (\ref{heisenberg_rho}) will tend to
\begin{equation*}
t(u)\rightarrow\exp(\xi^+ +\xi^- +\eta N-\eta\l/2)A_1+\exp(-\xi^+ -\xi^-+\eta\l/2-\eta N)D_1,
\end{equation*}
where\footnote{Note that the operators $A_1$, $B_1$, $C_1$ and $D_1$ differ by the absence of the shift $u\mapsto u-\eta/2$ and a scaling factor from the ones in the previous section.}
\begin{equation*}
R_{a\l}(u)... R_{a1}(u)=\begin{pmatrix}
                         A_1 & B_1 \\
                         C_1 & D_1
                        \end{pmatrix}.
\end{equation*}
 
By choosing $\gamma=\l/2-N-\eta^{-1}(\xi^+ +\xi^-)$ we can match it with the transfer matrix for the
closed chain, namely
\begin{equation*}
t(u)=\exp(-\eta\gamma)A_1+\exp(\eta\gamma)D_1={\rm tr}_a\left(\begin{pmatrix}
                                                        e^{-\eta\gamma} & 0 \\
                                                        0 & e^{\eta\gamma}
                                                        \end{pmatrix}_a R_{a\l}(u)...R_{a1}(u)\right).
\end{equation*}
Here again
\begin{equation*}
H=t^{-1}(0)t'(0)=\sum_{j=1}^{\l-1}H_{j(j+1)}+X_{\l}^{-1}H_{\l1}X_{\l}=\sum_{j=1}^{\l-1}H_{j(j+1)}+X_1 H_{\l1}X_1^{-1},
\end{equation*} 
where $H_{j(j+1)}=P_{j(j+1)}R'_{j(j+1)}(0)$ and $X=\begin{pmatrix}
                                                        e^{-\eta\gamma} & 0 \\
                                                        0 & e^{\eta\gamma}
                                                        \end{pmatrix}$.
  
In the rational limit (XXX model), the calculations are completely analogous to Section
\ref{sec:rational_limit}, so we omit the details.

As in Section \ref{sec:rational_limit}, we may summarise the analogous connections for the
Heisenberg model in the following diagram:
$$
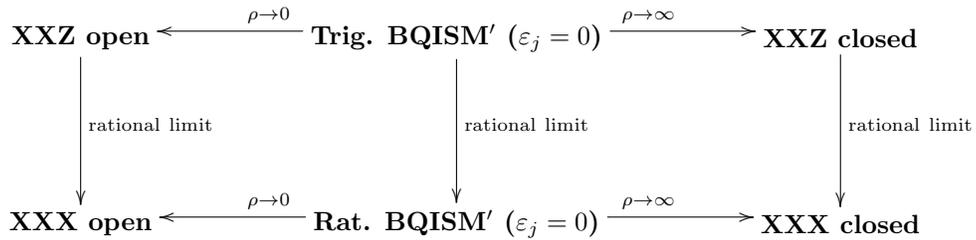
\xymatrix{
{\txt{\textbf{XXZ\ open}}}  \ar[dd]^{\rm rational\ limit} && 
\ar[ll]_{\rho\rightarrow 0} {\txt{\textbf{Trig. BQISM$^\prime$ ($\varepsilon_j=0$)}}} \ar[dd]^{\rm rational\ limit} \ar[rr]^{\rho\rightarrow\infty} &&
  {\txt{\textbf{XXZ closed}}}   \ar[dd]^{\rm rational\ limit}    \\
  && \\ 
{\txt{\textbf{XXX\ open}}} && \ar[ll]_{\rho\rightarrow 0} {\txt{\textbf{Rat. BQISM$^\prime$
($\varepsilon_j=0$)}}}  \ar[rr]^{\rho\rightarrow\infty} &&
{\txt{\textbf{XXX closed}}}               
}
$$
It is worth highlighting the fact that for the Heisenberg case, since we have set the parameters
$\varepsilon_j=0$, it is not possible to implement the variable change $\# 1$ discussed in the previous section.


\section{Quasi-classical limit and the spin-1/2 Richardson--Gaudin system}

Here we develop the main results of the current article. We investigate the quasi-classical limit of the system described in Section \ref{sec:BQISM}, which involves expanding all expressions in $\eta$ as $\eta\rightarrow 0$ and taking the first non-trivial term. 

In the quasi-classical limit, unlike the special case of the Heisenberg model above, we are able to implement variable change $\# 1$. Moreover, we gain the capability of implementing two additional variable changes. It is through these variable changes that we are able to make unexpected connections between various systems in the quasi-classical limit. We find that the following commutative
diagram, in contrast to those presented in Section \ref{sec:BQISM}, illustrates the connections we shall make in this section for the BAE and the conserved operators:
$$
\xymatrix{
{\txt{\textbf{Trig.\ BQISM}}}  \ar[dd]^{\rm rational\ limit}  \ar@<-1ex>[rr]_{\# 1}  && 
\ar[ll]_{\rho\rightarrow 0} {\txt{\textbf{Trig.\ BQISM$^\prime$}}} \ar[rr]^{\rho\rightarrow\infty} \ar[dd]^{\rm rational\ limit} &&
  {\txt{\textbf{Trig.\ QISM}}}   \ar@{-->}@<1ex>[ll]^{\# 3} \ar[dd]^{\rm rational\ limit} \\
  && \\ 
{\txt{\textbf{Rat.\ BQISM}}}  \ar@<-1ex>[rr]_{\# 1} \ar@{-->}@<1ex>[uu]^{\# 2} && 
\ar[ll]_{\rho\rightarrow 0} {\txt{\textbf{Rat.\ BQISM$^\prime$}}}  \ar[rr]^{\rho\rightarrow\infty}
\ar@{-->}@<1ex>[uu]^{\# 2} &&
{\txt{\textbf{Rat.\ QISM}}}               
}
$$
The connections that have been established previously still hold in the quasi-classical limit. Dashed arrows represent the connections that are yet to be established. In the diagram we adopt the
notation where $\# 1$ denotes variable change $\# 1$, $\# 2$ is used for variable change $\#2$ combined with some other operations, and $\# 3$ represents variable change $\# 3$ with different operations, all of which are specified explicitly in the text below.


\subsection{Bethe Ansatz Equations}\label{sec:bae}

We start by considering the BAE. Substituting the expressions (\ref{ad}) for $\tilde{a}(u)$ and $d(u)$ into the BAE (\ref{bae}) gives
\begin{equation}\label{bigBAE}
\begin{aligned}
&\frac{\sinh(\xi^+ +v_k+\rho/2+\eta/2)}{\sinh(\xi^+ -v_k-\rho/2+\eta/2)}\frac{\sinh(\xi^- +v_k+\rho/2+\eta/2)}{\sinh(\xi^- -v_k-\rho/2+\eta/2)}
\prod_{l=1}^{\mathcal{L}}\frac{\sinh(v_k-\eps_l-\eta/2)\sinh(v_k+\eps_l+\rho-\eta/2)}{\sinh(v_k-\eps_l+\eta/2)\sinh(v_k+\eps_l+\rho+\eta/2)}= \\
& \qquad\qquad =\prod_{j\neq k}^N\frac{\sinh(v_k-v_j-\eta)\sinh(v_k+v_j+\rho-\eta)}{\sinh(v_k-v_j+\eta)\sinh(v_k+v_j+\rho+\eta)}.
\end{aligned}
\end{equation}
If we set $\eta=0$ in (\ref{bigBAE}), the expression reduces to
\begin{equation}\label{baeqclim}
\frac{\sinh(\xi^- +v_k+\rho/2)\sinh(\xi^+ +v_k+\rho/2)}{\sinh(\xi^- -v_k-\rho/2)\sinh(\xi^+-v_k-\rho/2)}=1.
\end{equation}
Furthermore, we assume that $\xi^\pm$ depend on $\eta$ in such a way that (\ref{baeqclim}) holds as $\eta\rightarrow 0$.
We impose the following choice which is consistent with that property:
\begin{equation}\label{xi}
\xi^+=\xi+\eta\alpha,\ \ \xi^-=-\xi+\eta\beta.
\end{equation}
The expansion up to first order in $\eta$ for the right hand side of the BAE (\ref{bigBAE}) with (\ref{xi}) is given by  
\begin{equation*}
1-2\eta\sum_{j\neq k}^N\left(\coth(v_k-v_j)+\coth(v_k+v_j+\rho)\right).
\end{equation*}
Also, up to first order in $\eta$, the expansion of the left hand side of (\ref{bigBAE}) is
\begin{equation*}
1-\eta(\alpha+\beta+1)\left(\coth(v_k+\rho/2-\xi)+\coth(v_k+\rho/2+\xi)\right)-\eta\sum_{l=1}^\mathcal{L}\left(\coth(v_k-\eps_l)+\coth(v_k+\eps_l+\rho)\right).
\end{equation*}
Let us denote $\delta=-(\alpha+\beta+1)$. Then, in the limit as $\eta\rightarrow 0$, the BAE in the
case \textbf{Trig. BQISM$^\prime$} are given by
\begin{equation}\label{b_bae_trig'} 
\begin{aligned}
&\delta\left(\coth(v_k+\rho/2-\xi)+\coth(v_k+\rho/2+\xi)\right)+\sum_{l=1}^\mathcal{L}\left(\coth(v_k-\eps_l)+\coth(v_k+\eps_l+\rho)\right)=\\
& \qquad\qquad = 2\sum_{j\neq k}^N\left(\coth(v_k-v_j)+\coth(v_k+v_j+\rho)\right).
\end{aligned}
\end{equation}


\subsubsection{Variable change $\#1$}

It is a straightforward matter to see that \textbf{Trig. BQISM$^\prime$} (\ref{b_bae_trig'}) turns
into \textbf{Trig. BQISM} as $\rho\rightarrow 0$:
\begin{equation}\label{b_bae_trig}
  \begin{aligned}
  &\delta\left(\coth(v_k-\xi)+\coth(v_k+\xi)\right)+\sum_{l=1}^\mathcal{L}\left(\coth(v_k-\eps_l)+\coth(v_k+\eps_l)\right)=\\
  & \qquad\qquad = 2\sum_{j\neq k}^N\left(\coth(v_k-v_j)+\coth(v_k+v_j)\right).
  \end{aligned}
\end{equation}
Variable change $\# 1$ reverses this effect:
\begin{equation}
  v_k\mapsto v_k+\dfrac{\rho}{2},\ \eps_l\mapsto \eps_l+\dfrac{\rho}{2}.
\label{vc1}
\end{equation}


\subsubsection{Attenuated limit} 

As $\rho\rightarrow\infty$ \textbf{Trig. BQISM$^\prime$} (\ref{b_bae_trig'}) reduces to \textbf{Trig. QISM} in the quasi-classical limit:
\begin{equation*}
2\delta+\sum_{l=1}^\mathcal{L}\left(\coth(v_k-\eps_l)+1\right)=2\sum_{j\neq k}^N\left(\coth(v_k-v_j)+1\right),
\end{equation*}
or
\begin{equation}\label{bae_trig}
2\gamma+\sum_{l=1}^\mathcal{L}\coth(v_k-\eps_l)=2\sum_{j\neq k}^N\coth(v_k-v_j),
\end{equation}
where $\gamma=\delta+\l/2-(N-1)=-(\alpha+\beta+N-\l/2)$.\\

	
\subsubsection{Rational limit}

Introduce the rational parameter $\nu$ into \textbf{Trig. BQISM$^\prime$} (\ref{b_bae_trig'}):
\begin{equation*}
\begin{aligned}
&\delta\left(\coth\nu(v_k+\rho/2-\xi)+\coth\nu(v_k+\rho/2+\xi)\right)+\sum_{l=1}^\mathcal{L}\left(\coth\nu(v_k-\eps_l)+\coth\nu(v_k+\eps_l+\rho)\right)=\nonumber\\
& \qquad\qquad = 2\sum_{j\neq k}^N\left(\coth\nu(v_k-v_j)+\coth\nu(v_k+v_j+\rho)\right).
\end{aligned}
\end{equation*}
Multiplying by $\nu$ we obtain, since $\displaystyle \lim_{\nu\to 0}\dfrac{\nu\cosh(\nu x)}{\sinh(\nu x)}=\dfrac{1}{x}$, \textbf{Rat.\ BQISM$^\prime$} as $\nu\rightarrow 0$:
\begin{equation}\label{b_bae_rat'}
\frac{\delta}{(v_k+\rho/2)^2-\xi^2}+\sum_{l=1}^\mathcal{L}\frac{1}{(v_k+\rho/2)^2-(\eps_l+\rho/2)^2}=2\sum_{j\neq k}^N\frac{1}{(v_k+\rho/2)^2-(v_j+\rho/2)^2},
\end{equation}
which turns into \textbf{Rat.\ BQISM} as $\rho\rightarrow 0$:
\begin{equation}\label{b_bae_rat}
\frac{\delta}{v_k^2-\xi^2}+\sum_{l=1}^\mathcal{L}\frac{1}{v_k^2-\eps_l^2}=2\sum_{j\neq k}^N\frac{1}{v_k^2-v_j^2}.
\end{equation}

\subsubsection{Rational BQISM and trigonometric QISM equivalence} 

Make a change of variables $v_k\mapsto\ln y_k,\ \eps_l\mapsto\ln z_l$ in \textbf{Trig. QISM} (\ref{bae_trig}):
\begin{equation*}
2\delta+\sum_{l=1}^{\mathcal{L}}\left(\frac{y_k^2+z_l^2}{y_k^2-z_l^2}+1\right)=2\sum_{j\neq k}^N\left(\frac{y_k^2+y_j^2}{y_k^2-y_j^2}+1\right),
\end{equation*}
or
\begin{equation}\label{bae_trig''}
\delta+\sum_{l=1}^{\mathcal{L}}\frac{y_k^2}{y_k^2-z_l^2}=2\sum_{j\neq k}^N\frac{y_k^2}{y_k^2-y_j^2}.
\end{equation}
Note that \textbf{Rat. BQISM} (\ref{b_bae_rat}) turns into (\ref{bae_trig''}) under the following (invertible) variable change:
\begin{equation*}
v_k\mapsto\sqrt{y_k^2+\xi^2},\ \eps_l\mapsto\sqrt{z_l^2+\xi^2}.
\end{equation*}
Thus, \textbf{Trig. QISM} is equivalent to \textbf{Rat. BQISM} via the variable change from (\ref{bae_trig}) to (\ref{b_bae_rat}) given by
\begin{equation}\label{vcstar}
v_k\mapsto\ln\sqrt{v_k^2-\xi^2},\ \eps_l\mapsto\ln\sqrt{\eps_l^2-\xi^2},
\end{equation}
and its inverse
\begin{equation*}
v_k\mapsto\sqrt{\exp(2v_k)+\xi^2},\ \eps_l\mapsto\sqrt{\exp(2\eps_l)+\xi^2}
\end{equation*}
which obviously maps from (\ref{b_bae_rat}) to (\ref{bae_trig}).


\subsubsection{Variable change $\#2$}

It can be seen that we may transform from \textbf{Rat. BQISM} (\ref{b_bae_rat}) to \textbf{Trig.
BQISM} (\ref{b_bae_trig}) by a suitable variable change. Application of 
\begin{equation*}
v_k\mapsto\frac{y_k-y_k^{-1}}{2},\ \eps_l\mapsto\frac{z_l-z_l^{-1}}{2},\ \xi\mapsto\frac{\chi-\chi^{-1}}{2}
\end{equation*}
to \textbf{Rat.\ BQISM} (\ref{b_bae_rat}) gives
\begin{align*}
\delta\left(\frac{y_k^2+\chi^2}{y_k^2-\chi^2}+\frac{y_k^2\chi^2+1}{y_k^2\chi^2-1}\right)+
\sum_{l=1}^\mathcal{L}\left(\frac{y_k^2+z_l^2}{y_k^2-z_l^2}+\frac{y_k^2 z_l^2+1}{y_k^2 z_l^2-1}\right)=&
\,2\sum_{j\neq k}^N\left(\frac{y_k^2+y_j^2}{y_k^2-y_j^2}+\frac{y_k^2 y_j^2+1}{y_k^2 y_j^2-1}\right).
\end{align*}
Now, in order to transform this expression into \textbf{Trig. BQISM} (\ref{b_bae_trig}) we make a change of variables 
\begin{equation*}
y_k\mapsto \exp v_k,\ z_l\mapsto \exp\eps_l,\ \chi\mapsto \exp\xi. 
\end{equation*}
Thus, the mapping from \textbf{Rat. BQISM} (\ref{b_bae_rat}) to \textbf{Trig. BQISM} (\ref{b_bae_trig}) is a composition
\begin{equation}
\begin{aligned}
v_k&\mapsto
\sinh v_k,\\
\eps_l&\mapsto
\sinh\eps_l,\\
\xi&\mapsto
\sinh\xi.
\end{aligned}
\label{vc2norho}
\end{equation}
Analogously, including $\rho$ gives the mapping from \textbf{Rat. BQISM$^\prime$}
(\ref{b_bae_rat'}) to \textbf{Trig. BQISM$^\prime$} (\ref{b_bae_trig'}):
\begin{equation}
\begin{aligned}
v_k+\rho/2&\mapsto\sinh(v_k+\rho/2),\\
\eps_l+\rho/2&\mapsto\sinh(\eps_l+\rho/2),\\
\xi&\mapsto\sinh\xi.
\end{aligned}
\label{vc2}
\end{equation}
Generally, we refer to equations (\ref{vc2}) as the variable change $\#2$, and note that
(\ref{vc2norho}) is merely a specialisation of (\ref{vc2}) with $\rho=0$.


\subsubsection{Variable change $\# 3$}

Now, we define the variable change $\#3$ to be a composition comprising of operations defined so far: 

\ \\

\nid
\textbf{Trig. QISM} (\ref{bae_trig}) $\ \xrightarrow{\txt{(\ref{vcstar})}}\ $ \textbf{Rat. BQISM} (\ref{b_bae_rat}) 
$\ \xrightarrow{\txt{(\ref{vc2norho})}}\ $ \textbf{Trig. BQISM} (\ref{b_bae_trig}) $\
\xrightarrow{\txt{(\ref{vc1})}}\ $ \textbf{Trig. BQISM$^\prime$} (\ref{b_bae_trig'}).

\ \\

\nid This results in the variable change given by
\begin{equation}
\begin{aligned}
&v_k\mapsto\ln\sqrt{\sinh^2(v_k+\rho/2)-\sinh^2\xi},\\ 
&\eps_l\mapsto\ln\sqrt{\sinh^2(\eps_l+\rho/2)-\sinh^2\xi}.
\end{aligned}
\label{vc3}
\end{equation}
Equivalently, we may take

\ \\

\nid
\textbf{Trig. QISM} (\ref{bae_trig}) $\ \xrightarrow{\txt{(\ref{vcstar}) }}\ $ \textbf{Rat. BQISM} (\ref{b_bae_rat}) 
$\ \xrightarrow{\txt{(\ref{vc1})}}\ $ \textbf{Rat. BQISM$^\prime$} (\ref{b_bae_rat'}) $\
\xrightarrow{\txt{(\ref{vc2})}}\ $ \textbf{Trig. BQISM$^\prime$} (\ref{b_bae_trig'}),

\ \\

\nid which gives the same. We refer to the (\ref{vc3}) as variable change $\# 3$.


\subsubsection{Reduction to the rational, twisted-periodic case}

One can obtain \textbf{Rat. QISM} by taking the rational limit of \textbf{Trig. QISM} (\ref{bae_trig}). Introduce the
rational parameter $\nu$ into (\ref{bae_trig}):
\begin{equation*}
2\delta+\sum_{l=1}^\mathcal{L}\coth(\nu(v_k-\eps_l))=2\sum_{j\neq k}^N\coth(\nu(v_k-v_j)).
\end{equation*}
Then, denoting $\delta=\gamma/\nu$, multiply through by $\nu$ and consider $\nu\rightarrow 0$.
In such a case we obtain \textbf{Rat. QISM} in the quasi-classical limit:
\begin{equation}\label{bae_rat}
2\gamma+\sum_{l=1}^\mathcal{L}\frac{1}{v_k-\eps_l}=2\sum_{j\neq k}^N\frac{1}{v_k-v_j}.
\end{equation}
We can also obtain \textbf{Rat. QISM} (\ref{bae_rat}) by taking the attenuated limit from \textbf{Rat. BQISM$^\prime$} (\ref{b_bae_rat'}):
\begin{equation*}
\delta+\sum_{l=1}^\mathcal{L}\dfrac{v_k^2+\rho v_k+\rho^2/4-\xi^2}{v_k^2-\eps_l^2+\rho(v_k-\eps_l)}=
2\sum_{i\neq k}^N\dfrac{v_k^2+\rho v_k+\rho^2/4-\xi^2}{v_k^2-v_i^2+\rho(v_k-v_i)}.
\end{equation*}
Rescale the constant $\delta=\rho\gamma/2$, divide throughout by $\rho/4$ and consider $\rho\rightarrow\infty$. Then we obtain again \textbf{Rat. QISM} (\ref{bae_rat}):
\begin{equation*}
2\gamma+\sum_{l=1}^\mathcal{L}\frac{1}{v_k-\eps_l}=2\sum_{j\neq k}^N\frac{1}{v_k-v_j}.
\end{equation*}
Thus, we may summarise the connections made so far in the following diagram:
$$
\xymatrix{
{\txt{\textbf{Trig.\ BQISM}}}  \ar[dd]^{\rm rational\ limit}  \ar@<-1ex>[rr]_{\# 1}  && 
\ar[ll]_{\rho\rightarrow 0} {\txt{\textbf{Trig.\ BQISM$^\prime$}}} \ar[rr]^{\rho\rightarrow\infty} \ar[dd]^{\rm rational\ limit} &&
  {\txt{\textbf{Trig.\ QISM}}}   \ar@<1ex>[ll]^{\# 3} \ar[dd]^{\rm rational\ limit} \\
  && \\ 
{\txt{\textbf{Rat.\ BQISM}}}  \ar@<-1ex>[rr]_{\# 1} \ar@<1ex>[uu]^{\# 2} && 
\ar[ll]_{\rho\rightarrow 0} {\txt{\textbf{Rat.\ BQISM$^\prime$}}}  \ar[rr]^{\rho\rightarrow\infty}
\ar@<1ex>[uu]^{\# 2} &&
{\txt{\textbf{Rat.\ QISM}}}               
}
$$
It turns out that the limit labelled \textbf{Rat. QISM} is not equivalent to any of the other five nodes in the diagram above. This is deduced by knowledge of a particular solution of the BAE. For the BAE ({\ref{bae_trig''}), it was identified in \cite{ilsz09} that when $\delta=N-1$ there is a solution for which $y_k=0$ for all $k$. Results from \cite{rdo10} show that such a solution where all roots are equal does not exist for the BAE (\ref{bae_rat}). Consequently (\ref{bae_trig''}) and (\ref{bae_rat}) cannot be equivalent.     

The most unexpected aspect of the above calculations concerns the parameter $\xi$. Recall that this parameter arises in the expansion of the variables $\xi^\pm$, as given by (\ref{xi}), where $\xi^\pm$ are the free parametrising variables of the reflection matrices (\ref{K-matrices}). The above calculations show that $\xi$ is a spurious variable which can be removed by appropriate variable changes. In the next section we will show that it is also possible to remove the $\xi$-dependence from the conserved operators, but this requires an appropriate rescaling and basis transformation in conjunction with the variable changes.


\subsection{Conserved operators}\label{sec:im}

In the quasi-classical limit, the conserved operators $\tau_j$ are constructed as follows from the transfer matrix:
\begin{equation}
\lim_{u\rightarrow\eps_j}(u-\eps_j)\check{t}(u)=\eta^2\tau_j+o(\eta^2).
\label{TauDef}
\end{equation}
To calculate these conserved operators, we first set $\rho=0$, and impose the conditions (\ref{xi}) on
$\xi^\pm$ that appear in the reflection matrices given in equations (\ref{K-matrices}). 
Expanding $\check{K}^\pm(u)$ in $\eta$ as $\eta\rightarrow 0$ then gives
\begin{equation}\label{limK}
\check{K}^+(u)= \frac{1}{\sinh u}(K_1^+(u) +\eta K_2^+(u))+o(\eta),\ \ \check{K}^-(u)= \frac{1}{\sinh u}(K_1^-(u) +\eta K_2^-(u))+o(\eta),
\end{equation}
where we define 
\begin{equation*}
K_1^+(u)=\begin{pmatrix}
          \sinh(\xi+u) & 0 \\
          0 & \sinh(\xi-u)
          \end{pmatrix},\ \
K_2^+(u)=\begin{pmatrix}
         \left(\alpha+\frac{1}{2}\right)\cosh(\xi+u) & 0 \\
         0 & \left(\alpha-\frac{1}{2}\right)\cosh(\xi-u)
         \end{pmatrix},       
\end{equation*}
and
\begin{equation*}
K_1^-(u)=-\begin{pmatrix}
          \sinh(\xi-u) & 0 \\
          0 & \sinh(\xi+u)
          \end{pmatrix},\ \
K_2^-(u)=\begin{pmatrix}
         \left(\beta-\frac{1}{2}\right)\cosh(\xi-u) & 0 \\
         0 & \left(\beta+\frac{1}{2}\right)\cosh(\xi+u)
         \end{pmatrix}.      
\end{equation*}
It is easily verified that $\check{L}(u)$ given by (\ref{L-operator}) can be represented as follows:
\begin{equation}
\check{L}(u)=I+\frac{\eta}{\sinh u}r(u)+o(\eta),
\label{limr}
\end{equation}
where
\begin{equation*}
r(u)=\begin{pmatrix}
     S^z\cosh u  & S^- \\
     S^+ & -S^z\cosh u 
     \end{pmatrix}.
\end{equation*}
Here we have introduced the representation matrices of $su(2)$ corresponding to the fundamental
(i.e. two-dimensional) representation. Specifically, they are the matrices
$$
S^z = \frac12 \left( \begin{array}{cr} 1&0\\ 0&-1 \end{array} \right),\ \ 
S^+ = \left( \begin{array}{cr} 0&1\\ 0&0 \end{array} \right),\ \ 
S^- = \left( \begin{array}{cr} 0&0\\ 1&0 \end{array} \right)
$$
which satisfy the commutation relations
$$
[ S^z,S^\pm] = \pm S^\pm,\ \ [S^+,S^-]=2S^z.
$$
It is worth remarking that the connections that we make in the current article are only concerning this two-dimensional local Hilbert space. These are what we refer to as the spin-$1/2$ Richardson--Gaudin system.


Using the expressions of equations (\ref{limK}) and (\ref{limr}) above, we may take the expression (\ref{t-matrix-2}) for the transfer matrix and expand (\ref{TauDef}) explicitly as
\begin{equation*}
\begin{aligned}
\lim_{u\rightarrow\eps_j}(u-\eps_j)\check{t}(u)
&=\,\frac{1}{\sinh^2\eps_j}\eta \ 
{\rm tr}_a\Bigg[K_1^+(\varepsilon_j)r_{aj}(0)K_1^-(\varepsilon_j)
	+\eta\sum_{k>j}^\mathcal{L}\frac{K_1^+(\varepsilon_j)r_{ak}(\eps_j-\eps_k)r_{aj}(0)K_1^-(\varepsilon_j)}{\sinh(\eps_j-\eps_k)}+\\
	&+\eta\sum_{k<j}^\mathcal{L}\frac{K_1^+(\varepsilon_j)r_{aj}(0)r_{ak}(\eps_j-\eps_k)K_1^-(\varepsilon_j)}{\sinh(\eps_j-\eps_k)}+
  \eta K_2^+(\varepsilon_j)r_{aj}(0)K_1^-(\varepsilon_j) +\\
	&+\eta\sum_{k=1}^\mathcal{L}\frac{K_1^+(\varepsilon_j)r_{aj}(0)K_1^-(\varepsilon_j)r_{ak}(\eps_j+\eps_k)}{\sinh(\eps_j+\eps_k)}
+\eta K_1^+(\varepsilon_j)r_{aj}(0)K_2^-(\varepsilon_j)\Bigg]+o(\eta^2).
\end{aligned}
\end{equation*}
In the above expression each $K$-matrix acts on the auxiliary space, however we have suppressed the subscripts ``$a$'' for ease of notation. Finally, after computing the traces, we obtain
\begin{equation*}
\begin{aligned}
\tau_j= &\,\frac{1}{\sinh^2\eps_j}\Bigg[\sum_{k\neq j}^\mathcal{L}\frac{1}{\sinh(\eps_j-\eps_k)}\sinh(\eps_j+\xi)\sinh(\eps_j-\xi)\left(2\cosh(\eps_j-\eps_k)S_k^zS_j^z+S_k^-S_j^++S_k^+S_j^-\right)+\\
+&\sum_{k=1}^\mathcal{L}\frac{1}{\sinh(\eps_j+\eps_k)}\left(2\sinh(\eps_j+\xi)\sinh(\eps_j-\xi)\cosh(\eps_j+\eps_k)S_j^zS_k^z-\sinh^2(\eps_j+\xi)S_j^-S_k^+ -\sinh^2(\eps_j-\xi)S_j^+S_k^-\right)+\\
+&\left(\alpha\sinh(2\eps_j)-\frac{1}{2}\sinh(2\xi)\right)S_j^z+\left(\beta\sinh(2\eps_j)-\frac{1}{2}\sinh(2\xi)\right)S_j^z\Bigg].
\end{aligned}
\end{equation*}
We rescale and denote
${\tau}_j^{trig}=\dfrac{\sinh^2\eps_j}{\sinh(\eps_j+\xi)\sinh(\eps_j-\xi)}\tau_j,$ so that
\begin{equation}\label{b_im_trig}
\begin{aligned}
{\tau}_j^{trig}= &\sum_{k\neq j}^\mathcal{L}\frac{1}{\sinh(\eps_j-\eps_k)}\left(2\cosh(\eps_j-\eps_k)S_j^zS_k^z+S_j^+S_k^-+S_j^-S_k^+\right)+\\
+&\sum_{k=1}^\mathcal{L}\frac{1}{\sinh(\eps_j+\eps_k)}\left(2\cosh(\eps_j+\eps_k)S_j^zS_k^z
-\frac{\sinh(\eps_j-\xi)}{\sinh(\eps_j+\xi)}S_j^+S_k^--\frac{\sinh(\eps_j+\xi)}{\sinh(\eps_j-\xi)}S_j^-S_k^+\right)+\\
+&\,\frac{(\alpha+\beta)\sinh(2\eps_j)-\sinh(2\xi)}{\sinh(\eps_j+\xi)\sinh(\eps_j-\xi)}S_j^z.
\end{aligned}
\end{equation}
Thus, $\{{\tau}_j^{trig},\ j=1,...,\l\}$ are the mutually commuting conserved operators for \textbf{Trig. BQISM}.


\subsubsection{Variable change $\# 1$}
\label{vc1cons}

\nid The variable change $\# 1$ of equation (\ref{vc1}), particularly $\eps_j\mapsto\eps_j+{\rho}/{2}$, gives the conserved operators for \textbf{Trig. BQISM$^\prime$}:
\begin{equation}\label{b_im_trig'}
\begin{aligned}
{\tau}_j^{trig^\prime}=&\sum_{k\neq j}^\mathcal{L}\frac{1}{\sinh(\eps_j-\eps_k)}\left(2\cosh(\eps_j-\eps_k)S_j^zS_k^z+S_j^+S_k^-+S_j^-S_k^+\right)+\\
+&\sum_{k=1}^\mathcal{L}\frac{1}{\sinh(\eps_j+\eps_k+\rho)}\Big(2\cosh(\eps_j+\eps_k+\rho)S_j^zS_k^z-\\
-&\,\frac{\sinh(\eps_j+\rho/2-\xi)}{\sinh(\eps_j+\rho/2+\xi)}S_j^+S_k^--\frac{\sinh(\eps_j+\rho/2+\xi)}{\sinh(\eps_j+\rho/2-\xi)}S_j^-S_k^+ \Big)+\\
+&\,\frac{(\alpha+\beta)\sinh(2\eps_j+\rho)-\sinh(2\xi)}{\sinh(\eps_j+\rho/2+\xi)\sinh(\eps_j+\rho/2-\xi)}S_j^z.
\end{aligned}
\end{equation}

\subsubsection{Attenuated limit}

\nid Taking $\rho\rightarrow\infty$ in (\ref{b_im_trig'}) yields the conserved operators for \textbf{Trig. QISM}:
\begin{equation}\label{im_trig}
\begin{aligned}
{\tau}_j^{trig^\prime}\rightarrow\tau_j^{a.trig}=\sum_{k\neq j}^\mathcal{L}\frac{1}{\sinh(\eps_j-\eps_k)}\left(2\cosh(\eps_j-\eps_k)S_j^zS_k^z+S_j^+S_k^-+S_j^-S_k^+\right)-2\gamma S_j^z,
\end{aligned}
\end{equation}
where $\gamma=-(\alpha+\beta+N-\l/2)$, and the superscript``\textit{a.trig}'' refers to the attenuated limit of the trigonometric system.


\subsubsection{Rational limit}

\nid The rational limit of the conserved operators for \textbf{Trig. BQISM} (\ref{b_im_trig}) gives the conserved operators for \textbf{Rat. BQISM}:
\begin{equation}\label{b_im_rat}
\begin{aligned}
{\tau}_j^{rat}= &\sum_{k\neq j}^\mathcal{L}\frac{1}{\eps_j-\eps_k}\left(2S_j^zS_k^z+S_j^+S_k^-+S_j^-S_k^+\right)+\\
+&\sum_{k=1}^\mathcal{L}\frac{1}{\eps_j+\eps_k}\left(2S_j^zS_k^z
-\frac{\eps_j-\xi}{\eps_j+\xi}S_j^+S_k^--\frac{\eps_j+\xi}{\eps_j-\xi}S_j^-S_k^+ \right)+\\
+&\,\frac{2(\alpha+\beta)\eps_j-2\xi}{\eps_j^2-\xi^2}S_j^z.
\end{aligned}
\end{equation}  
We rewrite this expression as
\begin{equation*}
\begin{aligned}
{\tau}_j^{rat}=&\,4\sum_{k\neq j}^\l\frac{\eps_j}{\eps_j^2-\eps_k^2}S_j^zS_k^z+
\sum_{k\neq j}^\l\left(\frac{1}{\eps_j-\eps_k}-\frac{1}{\eps_j+\eps_k}\frac{\eps_j-\xi}{\eps_j+\xi}\right)S_j^+S_k^-+\\
+&\sum_{k\neq j}^\l\left(\frac{1}{\eps_j-\eps_k}-\frac{1}{\eps_j+\eps_k}\frac{\eps_j+\xi}{\eps_j-\xi}\right)S_j^-S_k^++\frac{I}{4\eps_j}
-\frac{1}{2\eps_j}\frac{\eps_j-\xi}{\eps_j+\xi}S_j^+S_j^--\frac{1}{2\eps_j}\frac{\eps_j+\xi}{\eps_j-\xi}S_j^-S_j^++\\
+&\,\frac{2(\alpha+\beta)\eps_j}{\eps_j^2-\xi^2}S_j^z-\frac{2\xi}{\eps_j^2-\xi^2}S_j^z.
\end{aligned}
\end{equation*} 
Using $S^+S^-={I}/{2}+S^z,\ \ S^-S^+={I}/{2}-S^z$ we obtain, after simplification and rescaling by $\eps_j$:
\begin{equation}\label{b_im_rat*}
\begin{aligned}
\eps_j{\tau}_j^{rat}=&\,4\sum_{k\neq j}^\l\frac{\eps_j^2}{\eps_j^2-\eps_k^2}S_j^zS_k^z+
2\sum_{k\neq j}^\l\frac{\eps_j^2}{\eps_j^2-\eps_k^2}\left(\frac{\eps_k+\xi}{\eps_j+\xi}S_j^+S_k^-+\frac{\eps_k-\xi}{\eps_j-\xi}S_j^-S_k^+\right)+\\
+&\,\frac{I}{4}-\frac{\eps_j^2+\xi^2}{\eps_j^2-\xi^2}\frac{I}{2}+\frac{2(\alpha+\beta)\eps_j^2}{\eps_j^2-\xi^2}S_j^z.
\end{aligned}
\end{equation}   


\subsubsection{Rational BQISM and trigonometric QISM equivalence}
\label{vcstarcons}

\nid Separating the terms with $\xi$ from the rest in equation (\ref{b_im_rat*}) we obtain the following equivalent expression:
\begin{equation*}
\begin{aligned}
\eps_j{\tau}_j^{rat}=&\,2\sum_{k\neq j}^\l\frac{\eps_j^2+\eps_k^2}{\eps_j^2-\eps_k^2}S_j^zS_k^z+
2\sum_{k\neq j}^\l\frac{\eps_j\eps_k}{\eps_j^2-\eps_k^2}\left(S_j^+S_k^- +S_j^-S_k^+\right)+2\left(\alpha+\beta+N-\frac{\l}{2}\right)S_j^z-\frac{3I}{4}+\\
+&\,2\sum_{k\neq j}^\l\frac{\eps_j}{\eps_j+\eps_k}\left(\frac{\xi}{\eps_j+\xi}S_j^+S_k^- -\frac{\xi}{\eps_j-\xi}S_j^-S_k^+\right)+
2\frac{\xi^2}{\eps_j^2-\xi^2}\left((\alpha+\beta)S_j^z-\frac{I}{2}\right).
\end{aligned}
\end{equation*} 
Now it is seen that if we set $\xi=0$ we obtain 
\begin{equation*}
\left.\eps_j{\tau}_j^{rat}\right|_{\xi=0}=\,2\sum_{k\neq j}^\l\frac{\eps_j^2+\eps_k^2}{\eps_j^2-\eps_k^2}S_j^zS_k^z+
2\sum_{k\neq j}^\l\frac{\eps_j\eps_k}{\eps_j^2-\eps_k^2}\left(S_j^+S_k^- +S_j^-S_k^+\right)+2\left(\alpha+\beta+N-\frac{\l}{2}\right)S_j^z-\frac{3I}{4}.
\end{equation*}  
The variable change $\eps_j\mapsto \exp\eps_j$ gives \textbf{Trig. QISM} (\ref{im_trig})  with $\gamma=-(\alpha+\beta+N-\l/2)$ (up to a constant term $-3I/4$):
\begin{equation*}
\left.\eps_j{\tau}_j^{rat}\right|_{\xi=0}\rightarrow \,2\sum_{k\neq j}^\l\coth(\eps_j-\eps_k)S_j^zS_k^z+
\sum_{k\neq j}^\l\frac{1}{\sinh(\eps_j-\eps_k)}\left(S_j^+S_k^- +S_j^-S_k^+\right)+2\left(\alpha+\beta+N-\frac{\l}{2}\right)S_j^z-\frac{3I}{4}.
\end{equation*}  
Now let us start with \textbf{Trig. QISM} (\ref{im_trig}) (with a change of variables $\eps_j=\ln z_j$):
\begin{equation*}
\tau_j^{(1)}=2\sum_{k\neq j}^\mathcal{L}\frac{z_j^2+z_k^2}{z_j^2-z_k^2}S_j^z S_k^z+2\sum_{k\neq j}^\mathcal{L}\frac{z_j z_k}{z_j^2-z_k^2}\left(S_j^+S_k^-+ S_j^-S_k^+\right)-2\gamma S_j^z.
\end{equation*}
Using $\dfrac{z_j^2+z_k^2}{z_j^2-z_k^2}=\dfrac{2z_j^2}{z_j^2-z_k^2}-1$ we obtain
\begin{equation*}
\tau_j^{(1)}=-2\sum_{k\neq j}^\l S_j^z S_k^z+4\sum_{k\neq j}^\l\frac{z_j^2}{z_j^2-z_k^2}S_j^z S_k^z+
2\sum_{k\neq j}^\mathcal{L}\frac{z_j z_k}{z_j^2-z_k^2}\left(S_j^+S_k^- + S_j^-S_k^+\right)-2\gamma S_j^z.
\end{equation*}
Furthermore, since 
\begin{equation*}
2\sum_{k\neq j}^\l S_j^z S_k^z=2\left(N-\frac{\l}{2}-S_j^z\right)S_j^z=2\left(N-\frac{\l}{2}\right)S_j^z-2(S_j^z)^2
\end{equation*}
and $(S^z)^2={I}/{4}$ for the spin-1/2 representation, we obtain
\begin{equation*}
\tau_j^{(1)}=4\sum_{k\neq j}^\l\frac{z_j^2}{z_j^2-z_k^2}S_j^z S_k^z+2\sum_{k\neq j}^\l\frac{z_k z_j}{z_j^2-z_k^2}\left(S_j^+S_k^- + S_j^-S_k^+\right)-
2\left(\gamma+N-\frac{\l}{2}\right) S_j^z+\frac{I}{2}.
\end{equation*}
A change of variable $z_j\mapsto\sqrt{\eps_j^2-\xi^2}$ gives the following conserved operators:
\begin{equation*}
\tau_j^{(2)}=4\sum_{k\neq j}^\l\frac{\eps_j^2-\xi^2}{\eps_j^2-\eps_k^2}S_j^z S_k^z+
2\sum_{k\neq j}^\mathcal{L}\frac{\sqrt{\eps_j^2-\xi^2}\sqrt{\eps_k^2-\xi^2}}{\eps_j^2-\eps_k^2}\left(S_j^+S_k^- + S_j^-S_k^+\right)-2\left(\gamma+N-\frac{\l}{2}\right) S_j^z+\frac{I}{2}.
\end{equation*}
Note that up to this point, all we have done is apply the change of variables given in
(\ref{vcstar}) on the $\varepsilon_j$. We further
rescale each conserved operator $\tau_j^{(2)}$ by the factor $\dfrac{\eps_j^2}{\eps_j^2-\xi^2}$:
\begin{equation*}
\tau_j^{(3)}=4\sum_{k\neq j}^\l\frac{\eps_j^2}{\eps_j^2-\eps_k^2}S_j^z S_k^z+
2\sum_{k\neq j}^\mathcal{L}\frac{\eps_j^2\sqrt{\eps_k^2-\xi^2}}{(\eps_j^2-\eps_k^2)\sqrt{\eps_j^2-\xi^2}}\left(S_j^+S_k^- + S_j^-S_k^+\right)-
2\left(\gamma+N-\frac{\l}{2}\right)\frac{\eps_j^2}{\eps_j^2-\xi^2} S_j^z+\frac{\eps_j^2}{\eps_j^2-\xi^2}\frac{I}{2}.
\end{equation*}
Consider a local transformation on the $j$th space in the tensor product
\begin{equation*}
 U_j=\text{diag}\left(\sqrt{\frac{\eps_j-\xi}{\eps_j+\xi}},1\right).
\end{equation*}
Under these transformations we have
\begin{equation*}
\begin{aligned}
U_j S_j^z U_j^{-1}=&\ S_j^z,\\
U_j S_j^+ U_j^{-1}=&\ \sqrt{\frac{\eps_j-\xi}{\eps_j+\xi}} S_j^+,\\
U_j S_j^- U_j^{-1}=&\ \sqrt{\frac{\eps_j+\xi}{\eps_j-\xi}}S_j^-. 
\end{aligned}
\end{equation*}
Under the global transformation $U=U_1 U_2...U_\l$ we find
\begin{equation*}
\begin{aligned}
\tau_j^{(4)}=&\ U \tau_j^{(3)} U^{-1}=\\
=&\,4\sum_{k\neq j}^\l\frac{\eps_j^2}{\eps_j^2-\eps_k^2}S_j^z S_k^z+
2\sum_{k\neq j}^\l\frac{\eps_j^2}{\eps_j^2-\eps_k^2}\left(\frac{\eps_k+\xi}{\eps_j+\xi}S_j^+S_k^- +\frac{\eps_k-\xi}{\eps_j-\xi}S_j^-S_k^+\right)-\\
-&\,2\left(\gamma+N-\frac{\l}{2}\right)\frac{\eps_j^2}{\eps_j^2-\xi^2} S_j^z+\frac{\eps_j^2}{\eps_j^2-\xi^2}\frac{I}{2}.
\end{aligned}
\end{equation*}
Note that these are the same as $\eps_j{\tau}_j^{rat}$ \textbf{Rat. BQISM} (\ref{b_im_rat*}), up to the constant term, taking into account that $\gamma=-(\alpha+\beta+N-\l/2)$.
Thus, we have
\begin{equation*}
\tau_j^{(4)}-\eps_j{\tau}_j^{rat}=\frac{\eps_j^2}{\eps_j^2-\xi^2}\frac{I}{2}+\frac{\eps_j^2+\xi^2}{\eps_j^2-\xi^2}\frac{I}{2}-\frac{I}{4}.
\end{equation*}
Finally, we can obtain
\begin{equation*}
{\tau}_j^{rat}=\frac{1}{\eps_j}\left(\tau_j^{(4)}-\frac{\eps_j^2}{\eps_j^2-\xi^2}\frac{I}{2}-\frac{\eps_j^2+\xi^2}{\eps_j^2-\xi^2}\frac{I}{2}+\frac{I}{4}\right).
\end{equation*}
 

\subsubsection{Variable change $\# 2$, rescaling, and a basis transformation} 
\label{vc2cons}

\nid Our goal now is to demonstrate how to transform \textbf{Rat. BQISM} (\ref{b_im_rat}) back into \textbf{Trig. BQISM} (\ref{b_im_trig}). 
First of all, we make a change of variables $\eps_j=\ln z_j,\ \xi=\ln\chi$ 
in \textbf{Trig. BQISM} (\ref{b_im_trig}):
\begin{equation*}
\begin{aligned}
{\tau}_j^{trig}=&\sum_{k\neq j}^\l\left(2\frac{z_j^2+z_k^2}{z_j^2-z_k^2}S_j^zS_k^z+\frac{2z_j z_k}{z_j^2-z_k^2}\left(S_j^+S_k^-+S_j^-S_k^+\right)\right)+\\
+&\sum_{k=1}^\l\left(2\frac{z_j^2 z_k^2+1}{z_j^2 z_k^2-1}S_j^zS_k^z-
\frac{2z_j z_k}{z_j^2z_k^2-1}\left(\frac{z_j^2-\chi^2}{\chi^2 z_j^2-1}S_j^+S_k^-+\frac{\chi^2 z_j^2-1}{z_j^2-\chi^2}S_j^-S_k^+ \right)\right)+\\
+&\,2\frac{(\alpha+\beta)\chi^2(z_j^4-1)-z_j^2(\chi^4-1)}{(\chi^2 z_j^2-1)(z_j^2-\chi^2)}S_j^z=\\
=&\,2\sum_{k\neq j}^\l\left(\frac{z_j^2+z_k^2}{z_j^2-z_k^2}+\frac{z_j^2 z_k^2+1}{z_j^2 z_k^2-1}\right)S_j^zS_k^z+\\
+&\,2\sum_{k\neq j}^\l\left[\left(\frac{z_j z_k}{z_j^2-z_k^2}-\frac{z_j z_k}{z_j^2z_k^2-1}\frac{z_j^2-\chi^2}{\chi^2 z_j^2-1}\right)S_j^+S_k^-+
\left(\frac{z_j z_k}{z_j^2-z_k^2}-\frac{z_j z_k}{z_j^2z_k^2-1}\frac{\chi^2 z_j^2-1}{z_j^2-\chi^2}\right)S_j^-S_k^+\right]+\\
+&\,\frac{z_j^4+1}{z_j^4-1}\frac{I}{2}-\frac{2 z_j^2}{z_j^4-1}\frac{z_j^2-\chi^2}{\chi^2 z_j^2-1}S_j^+S_j^--\frac{2 z_j^2}{z_j^4-1}\frac{\chi^2 z_j^2-1}{z_j^2-\chi^2}S_j^-S_j^++\\
+&\,\frac{2(\alpha+\beta)\chi^2(z_j^4-1)}{(\chi^2 z_j^2-1)(z_j^2-\chi^2)}S_j^z-\frac{2z_j^2(\chi^4-1)}{(\chi^2 z_j^2-1)(z_j^2-\chi^2)}S_j^z.
\end{aligned}
\end{equation*}
Using $S^+S^-={I}/{2}+S^z,\ \ S^-S^+={I}/{2}-S^z$ and simplifying we obtain
\begin{equation}\label{b_im_trig*}
\begin{aligned}
{\tau}_j^{trig}=&\,2\sum_{k\neq j}^\l\left(\frac{z_j^2+z_k^2}{z_j^2-z_k^2}+\frac{z_j^2 z_k^2+1}{z_j^2 z_k^2-1}\right)S_j^zS_k^z+\\
+&\,2\sum_{k\neq j}^\l\frac{z_j z_k(z_j^4-1)}{(z_j^2-z_k^2)(z_j^2 z_k^2-1)}\left[\frac{z_k^2\chi^2-1}{z_j^2\chi^2-1}S_j^+S_k^- + \frac{z_k^2-\chi^2}{z_j^2-\chi^2}S_j^-S_k^+\right]+\\
+&\,\frac{z_j^4+1}{z_j^4-1}\frac{I}{2}-\frac{z_j^2}{z_j^4-1}\left(\frac{z_j^2-\chi^2}{\chi^2 z_j^2-1}+\frac{\chi^2 z_j^2-1}{z_j^2-\chi^2}\right)I+
\frac{2(\alpha+\beta)\chi^2(z_j^4-1)}{(\chi^2 z_j^2-1)(z_j^2-\chi^2)}S_j^z.
\end{aligned}
\end{equation}
We begin with the the form (\ref{b_im_rat*}) of \textbf{Rat. BQISM}, multiplied by $\eps_j$:
\begin{equation*}
\begin{aligned}
\tilde{\tau}_j^{(1)}=\eps_j{\tau}_j^{rat}= &\sum_{k\neq j}^\l\frac{4\eps_j^2}{\eps_j^2-\eps_k^2}S_j^zS_k^z+
\sum_{k\neq j}^\l\frac{2\eps_j^2}{\eps_j^2-\eps_k^2}\left(\frac{\eps_k+\xi}{\eps_j+\xi}S_j^+S_k^-+\frac{\eps_k-\xi}{\eps_j-\xi}S_j^-S_k^+\right)+\\
+&\,\frac{I}{4}-\frac{\eps_j^2+\xi^2}{\eps_j^2-\xi^2}\frac{I}{2}+\frac{2(\alpha+\beta)\eps_j^2}{\eps_j^2-\xi^2}S_j^z.
\end{aligned}
\end{equation*} 
Now, make a change of variables $\eps_j\mapsto\dfrac{z_j-z_j^{-1}}{2},\ \ \xi\mapsto\dfrac{\chi-\chi^{-1}}{2}$:
\begin{equation*}
\begin{aligned}
\tilde{\tau}_j^{(2)}= &\sum_{k\neq j}^\l\frac{4(z_j-z_j^{-1})^2}{(z_j-z_j^{-1})^2-(z_k-z_k^{-1})^2}S_j^zS_k^z+\\
+&\sum_{k\neq j}^\l\frac{2(z_j-z_j^{-1})^2}{(z_j-z_j^{-1})^2-(z_k-z_k^{-1})^2}
\left(\frac{z_k-z_k^{-1}+\chi-\chi^{-1}}{z_j-z_j^{-1}+\chi-\chi^{-1}}S_j^+S_k^-
+\frac{z_k-z_k^{-1}-\chi+\chi^{-1}}{z_j-z_j^{-1}-\chi+\chi^{-1}}S_j^-S_k^+\right)+\\
+&\,\frac{I}{4}-\frac{(z_j-z_j^{-1})^2+(\chi-\chi^{-1})^2}{(z_j-z_j^{-1})^2-(\chi-\chi^{-1})^2}\frac{I}{2}+\frac{2(\alpha+\beta)(z_j-z_j^{-1})^2}{(z_j-z_j^{-1})^2-(\chi-\chi^{-1})^2}S_j^z.
\end{aligned}
\end{equation*} 
Then, rescale by $\dfrac{z_j^2-z_j^{-2}}{(z_j-z_j^{-1})^2}$:
\begin{equation*}
\begin{aligned}
\tilde{\tau}_j^{(3)}=&\sum_{k\neq j}^\l\frac{4(z_j^2-z_j^{-2})}{(z_j-z_j^{-1})^2-(z_k-z_k^{-1})^2}S_j^zS_k^z+\\
+&\sum_{k\neq j}^\l\frac{2(z_j^2-z_j^{-2})}{(z_j-z_j^{-1})^2-(z_k-z_k^{-1})^2}
\left(\frac{z_k-z_k^{-1}+\chi-\chi^{-1}}{z_j-z_j^{-1}+\chi-\chi^{-1}}S_j^+S_k^-
+\frac{z_k-z_k^{-1}-\chi+\chi^{-1}}{z_j-z_j^{-1}-\chi+\chi^{-1}}S_j^-S_k^+\right)+\\
+&\,\frac{z_j^2-z_j^{-2}}{(z_j-z_j^{-1})^2}\frac{I}{4}-\frac{(z_j-z_j^{-1})^2+(\chi-\chi^{-1})^2}{(z_j-z_j^{-1})^2-(\chi-\chi^{-1})^2}\dfrac{z_j^2-z_j^{-2}}{(z_j-z_j^{-1})^2}\frac{I}{2}+
\frac{2(\alpha+\beta)(z_j^2-z_j^{-2})}{(z_j-z_j^{-1})^2-(\chi-\chi^{-1})^2}S_j^z.
\end{aligned}
\end{equation*} 
Using the identities
\begin{equation*}
\begin{aligned}
\frac{2(z_j^2-z_j^{-2})}{(z_j-z_j^{-1})^2-(z_k-z_k^{-1})^2}=&\,\frac{z_j^2+z_k^2}{z_j^2-z_k^2}+\frac{z_j^2 z_k^2+1}{z_j^2 z_k^2-1}=\frac{2z_k^2(z_j^4-1)}{(z_j^2-z_k^2)(z_j^2 z_k^2-1)},\\
\frac{2(z_j^2-\chi^{-2})}{(z_j-z_j^{-1})^2-(\chi-\chi^{-1})^2}=&\,\frac{z_j^2+\chi^2}{z_j^2-\chi^2}+\frac{z_j^2 \chi^2+1}{z_j^2 \chi^2-1}=\frac{2\chi^2(z_j^4-1)}{(z_j^2-\chi^2)(z_j^2 \chi^2-1)},
\end{aligned}
\end{equation*} 
we obtain
\begin{equation*}
\begin{aligned}
\tilde{\tau}_j^{(3)}=&\,2\sum_{k\neq j}^\l\left(\frac{z_j^2+z_k^2}{z_j^2-z_k^2}+\frac{z_j^2 z_k^2+1}{z_j^2 z_k^2-1}\right)S_j^zS_k^z+\\
+&\,2\sum_{k\neq j}^\l\frac{z_k^2(z_j^4-1)}{(z_j^2-z_k^2)(z_j^2 z_k^2-1)}
\left(\frac{z_k-z_k^{-1}+\chi-\chi^{-1}}{z_j-z_j^{-1}+\chi-\chi^{-1}}S_j^+S_k^-
+\frac{z_k-z_k^{-1}-\chi+\chi^{-1}}{z_j-z_j^{-1}-\chi+\chi^{-1}}S_j^-S_k^+\right)+\\
+&\,\frac{z_j^2-z_j^{-2}}{(z_j-z_j^{-1})^2}\frac{I}{4}-
\frac{(z_j-z_j^{-1})^2+(\chi-\chi^{-1})^2}{(z_j-z_j^{-1})^2}\frac{\chi^2(z_j^4-1)}{(z_j^2-\chi^2)(z_j^2 \chi^2-1)}\frac{I}{2}+
2(\alpha+\beta)\frac{\chi^2(z_j^4-1)}{(z_j^2-\chi^2)(z_j^2 \chi^2-1)}S_j^z.
\end{aligned}
\end{equation*} 
Now we see that the first term already matches with the first term of (\ref{b_im_trig*}). To match the second term we need to make a basis transformation of the type $U=U_1 U_2...U_\l$, where
\begin{equation*}
 U_j=\text{diag}(x_j,1)
\end{equation*}
with
\begin{equation*}
x_j=\frac{z_j(z_j-z_j^{-1}+\chi-\chi^{-1})}{z_j^2\chi^2-1}.
\end{equation*}
Finally, we have
\begin{equation*}
\begin{aligned}
\tilde{\tau}_j^{(4)}=&\ U \tilde{\tau}_j^{(3)} U^{-1}=\\
=&\,2\sum_{k\neq j}^\l\left(\frac{z_j^2+z_k^2}{z_j^2-z_k^2}+\frac{z_j^2 z_k^2+1}{z_j^2 z_k^2-1}\right)S_j^zS_k^z+\\
+&\,2\sum_{k\neq j}^\l\frac{z_j z_k(z_j^4-1)}{(z_j^2-z_k^2)(z_j^2 z_k^2-1)}\left[\frac{z_k^2\chi^2-1}{z_j^2\chi^2-1}S_j^+S_k^- + \frac{z_k^2-\chi^2}{z_j^2-\chi^2}S_j^-S_k^+\right]+\\
+&\,\frac{z_j^2-z_j^{-2}}{(z_j-z_j^{-1})^2}\frac{I}{4}-
\frac{(z_j-z_j^{-1})^2+(\chi-\chi^{-1})^2}{(z_j-z_j^{-1})^2}\frac{\chi^2(z_j^4-1)}{(z_j^2-\chi^2)(z_j^2 \chi^2-1)}\frac{I}{2}+\\
+&\frac{2(\alpha+\beta)\chi^2(z_j^4-1)}{(\chi^2 z_j^2-1)(z_j^2-\chi^2)}S_j^z,
\end{aligned}
\end{equation*}
which is the same as ${\tau}_j^{trig}$ (\ref{b_im_trig*}) up to the constant term:
\begin{equation*}
\begin{aligned}
\tilde{\tau}_j^{(4)}-{\tau}_j^{trig}=&\frac{z_j^2-z_j^{-2}}{(z_j-z_j^{-1})^2}\frac{I}{4}-
\frac{(z_j-z_j^{-1})^2+(\chi-\chi^{-1})^2}{(z_j-z_j^{-1})^2}\frac{\chi^2(z_j^4-1)}{(z_j^2-\chi^2)(z_j^2 \chi^2-1)}\frac{I}{2}-\\
-&\,\frac{z_j^4+1}{z_j^4-1}\frac{I}{2}+\frac{z_j^2}{z_j^4-1}\left(\frac{z_j^2-\chi^2}{\chi^2 z_j^2-1}+\frac{\chi^2 z_j^2-1}{z_j^2-\chi^2}\right)I.
\end{aligned}
\end{equation*}


\subsubsection{Variable change $\# 3$, rescaling, and a basis transformation}

As in the case of the BAE, variable change $\# 3$ is defined as the composition which leads to (\ref{vc3}). Combined with the appropriate composition of basis transformations and rescalings described above, this leads to the following mappings for the conserved operators:

\ \\

\nid
\textbf{Trig. QISM} (\ref{im_trig}) $\ \xrightarrow{\txt{\ref{vcstarcons}}}\ $  \textbf{Rat. BQISM} (\ref{b_im_rat}) 
$\ \xrightarrow{\txt{\ref{vc2cons}}}\ $ \textbf{Trig. BQISM} (\ref{b_im_trig}) $\ \xrightarrow{\txt{\ref{vc1cons}}}\ $ \textbf{Trig. BQISM$^\prime$} (\ref{b_im_trig'}),

\ \\
where the arrow labels refer to the subsections where the corresponding operations are described.

\subsubsection{Reduction to the rational, twisted-periodic case}

\nid In the rational limit of \textbf{Trig. QISM} (\ref{im_trig}) we obtain the following conserved operators:
\begin{equation}\label{im_rat}
\tau_j^{a.rat}=-2\gamma S_j^z+\sum_{k\neq j}^\mathcal{L}\frac{2 S_j^z S_k^z+S_j^+S_k^- + S_j^-S_k^+}{\eps_j-\eps_k}.
\end{equation}
We can also obtain them via the attenuated limit from \textbf{Rat. BQISM} (\ref{b_im_rat}). First
introduce $\rho$ by the variable change $\# 1$:
\begin{equation}\label{b_im_rat'}
\begin{aligned}
{\tau}_j^{rat^\prime}= &\sum_{k\neq j}^\mathcal{L}\frac{1}{\eps_j-\eps_k}\left(2S_j^zS_k^z+S_j^+S_k^-+S_j^-S_k^+\right)+\\
+&\sum_{k=1}^\mathcal{L}\frac{1}{\eps_j+\eps_k+\rho}\left(2S_j^zS_k^z
-\frac{\eps_j+\rho/2-\xi}{\eps_j+\rho/2+\xi}S_j^+S_k^--\frac{\eps_j+\rho/2+\xi}{\eps_j+\rho/2-\xi}S_j^-S_k^+ \right)+\\
+&\,\frac{2(\alpha+\beta)(\eps_j+\rho/2)-2\xi}{(\eps_j+\rho/2)^2-\xi^2}S_j^z.
\end{aligned}
\end{equation}  
Choose $(\alpha+\beta)=-\gamma\rho/2$. Then (\ref{b_im_rat'}) tends to (\ref{im_rat}) as $\rho\rightarrow\infty$.


\section{Conclusion}
In this work we have studied the spin-1/2 Richardson--Gaudin system as the quasi-classical limit of a formulation provided by a generalised BQISM. In this manner we uncovered some surprising features, viz. that  
the rational limit of the boundary trigonometric system is equivalent to the original boundary trigonometric system.  Additionally we found that the twisted-periodic and boundary constructions are equivalent in the trigonometric case, but not in the rational limit. One consequence of this finding is that for the spin-1/2 Richardson--Gaudin system the BQISM formalism does not extend the integrable structure beyond that provided by the QISM formalism. This is an unexpected result, in contrast to the Heisenberg model.

There are several directions for future studies. One is to investigate the analogous system obtained by implementing non-diagonal solutions of the reflection equations. Due to the breaking of $u(1)$ symmetry in this instance, there is the possibility to make connection with elliptic parametrisations. The construction of conserved operators for this case has previously been undertaken in \cite{yzg04}, and we have already initiated an analysis of this problem. Higher spin versions of the Richardson--Gaudin system is another option. The BQISM formulation of these systems appears in the work \cite{dahog02}. Whether a basis transformation exists to establish the equivalence between the  \textbf{Rat. BQISM} and \textbf{Trig. QISM} conserved operators in this case remains an open problem, but examination of the associated BAE in \cite{dahog02} is suggestive that it does exist. Models based on higher rank algebras are also worthy of investigation. In this regard, a systematic construction of conserved operators has been undertaken in \cite{s12,s13} which unifies previous particular case studies. Supersymmetric analogues, such as the $osp(1|2)$ Richardson--Gaudin system \cite{km03}, provide another avenue for future research.   

\subsection*{Acknowledgements} Inna Lukyanenko thanks Prof. Wen-Li Yang and his group at the Institute of Modern Physics, Northwest University, China, for their kind hospitality. Phillip Isaac is supported by the Australian Research Council through Discovery Project DP140101492. Jon Links and Inna Lukyanenko are supported by the Australian Research Council through Discovery Project DP110101414. Inna Lukyanenko is funded through an International Postgraduate Scholarship and a UQ International Scholarship.

\appendix

\section{Eigenvalues of the conserved operators}\label{sec:ev}


In this article we have shown, in the quasi-classical limit, the explicit connections between the
BAE and conserved operators
associated with the rational limit of the BQISM for Richardson-Gaudin systems, and the corresponding
twisted-periodic trigonometric systems. We can also verify analogous connections
between the eigenvalues of the conserved operators. While this necessarily follows from the equivalence of the conserved operators, it is useful as a consistency
check as well as having the potential to provide some alternative insights into the methods used.
The summary diagram for the BAE, with the same variable changes, also holds on the level of eigenvalue formulae.
 
The eigenvalues $\lambda_j$ in the quasi-classical limit are constructed from (\ref{lambda}) as follows (set $\rho=0$):
\begin{equation*}
\lim_{u\rightarrow\eps_j}(u-\eps_j)\check{\Lambda}(u)=\eta^2\lambda_j+o(\eta^2).
\end{equation*}
It gives the eigenvalues for \textbf{Trig. BQISM} up to a factor of 
$
\displaystyle{\frac{\sinh^2\eps_j}{\sinh(\eps_j+\xi)\sinh(\eps_j-\xi)} }
$
as follows: 
\begin{equation}\label{b_ev_trig}
\begin{aligned}
{\lambda}_j^{trig}=&\,\frac{\delta}{2}\left(\coth(\eps_j-\xi)+\coth(\eps_j+\xi)\right)+\frac{3}{2}\coth(2\eps_j)\,+\\
+&\,\frac{1}{2}\sum_{k\neq j}^\l\left(\coth(\eps_j-\eps_k)+\coth(\eps_j+\eps_k)\right)
-\sum_{i=1}^N\left(\coth(\eps_j-v_i)+\coth(\eps_j+v_i)\right),
\end{aligned}
\end{equation}
where $\delta=-(\alpha+\beta+1)$.
We can check that the constant terms agree. To do this, we need to check that the action of
${\tau}_j^{trig}$ on the state $\Omega$, where
$\Omega=\begin{pmatrix}
         0 \\
         1 
         \end{pmatrix}^{\o\l}$, 
is equal to the constant term in (\ref{b_ev_trig}). Namely, that 
\begin{equation*}
\begin{aligned}
{\tau}_j^{trig}\Omega &=\Bigg(\frac{1}{2}\sum_{k\neq j}^\mathcal{L}\left(\coth(\eps_j-\eps_k)+\coth(\eps_j+\eps_k)\right)+\frac{1}{2}\coth(2\eps_j)
-\frac{1}{\sinh(2\eps_j)}\frac{\sinh(\eps_j+\xi)}{\sinh(\eps_j-\xi)}\,-\\
&\quad\quad\quad -\,\frac{1}{2}\frac{(\alpha+\beta)\sinh(2\eps_j)}{\sinh(\eps_j+\xi)\sinh(\eps_j-\xi)}+\frac{1}{2}\frac{\sinh(2\xi)}{\sinh(\eps_j+\xi)\sinh(\eps_j-\xi)}\Bigg)\Omega=\\
&=\Bigg(-\frac{1}{2}(\alpha+\beta+1)\left(\coth(\eps_j-\xi)+\coth(\eps_j+\xi)\right)+\frac{3}{2}\coth(2\eps_j)
+\,\frac{1}{2}\sum_{k\neq j}^\mathcal{L}\left(\coth(\eps_j-\eps_k)+\coth(\eps_j+\eps_k)\right)\Bigg)\Omega.
\end{aligned}
\end{equation*}
Indeed, by making repeated use of the identity
\begin{equation*}
\sinh(x+y)=\sinh(x)\cosh(y)+\cosh(x)\sinh(y)
\end{equation*}
and other similar identities for hyperbolic functions, we may easily check that
   \begin{equation*}
   \coth(\eps_j-\xi)+\coth(\eps_j+\xi)=\frac{\sinh(2\eps_j)}{\sinh(\eps_j+\xi)\sinh(\eps_j-\xi)}
   \end{equation*}
and
   \begin{equation*}
   -\,\frac{1}{ \sinh(2\eps_j)}\frac{\sinh(\eps_j+\xi)}{\sinh(\eps_j-\xi)}
   +\frac{1}{2}\frac{\sinh(2\xi)}{\sinh(\eps_j+\xi)\sinh(\eps_j-\xi)}=
   \coth(2\eps_j)-\frac{1}{2}\frac{\sinh(2\eps_j)}{\sinh(\eps_j+\xi)\sinh(\eps_j-\xi)}.
   \end{equation*}
Therefore ${\tau}_j^{trig}\Omega = \lambda_j^{trig}\Omega$ with $\lambda_j^{trig}$ given by equation (\ref{b_ev_trig}).
  
\ \\
 
\nid \underline{ {\bf Variable change $\# 1$}}

\ \\

\nid We can obtain \textbf{Trig. BQISM$^\prime$} by applying the variable change $\# 1$ given in
(\ref{vc1}):
\begin{equation}\label{b_ev_trig'}
\begin{aligned}
{\lambda}_j^{trig^\prime}=&\,\frac{\delta}{2}\left(\coth(\eps_j+\rho/2-\xi)+\coth(\eps_j+\rho/2+\xi)\right)+\frac{3}{2}\coth(2\eps_j+\rho)\,+\\
+&\,\frac{1}{2}\sum_{k\neq j}^\l\left(\coth(\eps_j-\eps_k)+\coth(\eps_j+\eps_k+\rho)\right)
-\sum_{i=1}^N\left(\coth(\eps_j-v_i)+\coth(\eps_j+v_i+\rho)\right).
\end{aligned}
\end{equation}


\ \\

\nid \underline{ {\bf Attenuated limit} }

\ \\

\nid Now, as $\rho\rightarrow\infty$ in \textbf{Trig. BQISM$^\prime$} (\ref{b_ev_trig'}), we obtain \textbf{Trig. QISM}:
\begin{equation*}
{\lambda}_j^{trig^\prime}\rightarrow\lambda_j^{a.trig}=
\delta+\frac{3}{2}+\frac{1}{2}\sum_{k\neq j}^\l(\coth(\eps_j-\eps_k)+1)-\sum_{i=1}^N(\coth(\eps_j-v_i)+1),
\end{equation*}
or
\begin{equation}\label{ev_trig}
\lambda_j^{a.trig}=\gamma+\frac{1}{2}\sum_{k\neq j}^\l\coth(\eps_j-\eps_k)-\sum_{i=1}^N\coth(\eps_j-v_i),
\end{equation}
where $\gamma=-(\alpha+\beta+N-\l/2)$. 

\ \\

\nid \underline{ {\bf Rational limit}}

\ \\

\nid The rational limit of \textbf{Trig. BQISM} (\ref{b_ev_trig}) gives \textbf{Rat. BQISM}:
\begin{equation}\label{b_ev_rat}
{\lambda}_j^{rat}=\,\frac{\delta\eps_j}{\eps_j^2-\xi^2}+\frac{3}{4\eps_j}+\sum_{k\neq j}^\l\frac{\eps_j}{\eps_j^2-\eps_k^2}-
\sum_{i=1}^N\frac{2\eps_j}{\eps_j^2-v_i^2}.
\end{equation}
Or, multiplied by $\eps_j$:
\begin{equation}\label{b_ev_rat*}
\eps_j{\lambda}_j^{rat}=\,\frac{\delta\eps_j^2}{\eps_j^2-\xi^2}+\frac{3}{4}+\sum_{k\neq j}^\l\frac{\eps_j^2}{\eps_j^2-\eps_k^2}-
\sum_{i=1}^N\frac{2\eps_j^2}{\eps_j^2-v_i^2}.
\end{equation}

\ \\

\nid \underline{ {\bf Equivalence of the rational BQISM and the trigonometric QISM}}

\ \\

\nid Set $\xi=0$ in \textbf{Rat. BQISM} (\ref{b_ev_rat*}):
\begin{equation*}
\eps_j{\lambda}_j^{rat}|_{\xi=0}=\,\delta+\frac{3}{4}+\sum_{k\neq j}^\l\frac{\eps_j^2}{\eps_j^2-\eps_k^2}-\sum_{i=1}^N\frac{2\eps_j^2}{\eps_j^2-v_i^2}.
\end{equation*}
Using \ $\dfrac{\eps_j^2}{\eps_j^2-\eps_k^2}=\dfrac{1}{2}\left(\dfrac{\eps_j^2+\eps_k^2}{\eps_j^2-\eps_k^2}+1\right)$ we obtain
\begin{equation*}
\eps_j{\lambda}_j^{rat}|_{\xi=0}=\,\delta+\frac{3}{4}+\frac{(\l-1)}{2}-N+\frac{1}{2}\sum_{k\neq j}^\l\frac{\eps_j^2+\eps_k^2}{\eps_j^2-\eps_k^2}-\sum_{i=1}^N\frac{\eps_j^2+v_i^2}{\eps_j^2-v_i^2}.
\end{equation*}
Making a change of variables $\eps_j\mapsto \exp\eps_j$, we obtain \textbf{Trig. QISM} (\ref{ev_trig}) up to a constant term $-3/4$:
\begin{equation*}
\eps_j{\lambda}_j^{rat}|_{\xi=0}=\,-\left(\alpha+\beta+N-\frac{\l}{2}\right)-\frac{3}{4}+\frac{1}{2}\sum_{k\neq j}^\l\coth(\eps_j-\eps_k)-\sum_{i=1}^N\coth(\eps_j-v_i).
\end{equation*} 
Now, we want to turn \textbf{Trig. QISM} (\ref{ev_trig}) back into \textbf{Rat. BQISM} (\ref{b_ev_rat*}). 
We start with \textbf{Trig. QISM} (\ref{ev_trig}) (with a change of variables $\eps_j=\ln z_j,\ v_i=\ln y_i$)
\begin{equation*}
\begin{aligned}
\lambda^{(1)}=&\,\gamma+\frac{1}{2}\sum_{k\neq j}^\l\frac{z_j^2+z_k^2}{z_j^2-z_k^2}-\sum_{i=1}^N\frac{z_j^2+y_i^2}{z_j^2-y_i^2}=\\
=&\,\gamma+N-\frac{\l}{2}+\frac{1}{2}+\sum_{k\neq j}^\l \frac{z_j^2}{z_j^2-z_k^2}-2\sum_{i=1}^N\frac{z_j^2}{z_j^2-y_i^2}.
\end{aligned}
\end{equation*}
Make the change of variables 
\begin{equation*}
z_j\mapsto\sqrt{\eps_j^2-\xi^2},\ y_i\mapsto\sqrt{v_i^2-\xi^2}.
\end{equation*}
This gives 
\begin{equation*}
\lambda_j^{(2)}=\gamma+N-\frac{\l}{2}+\frac{1}{2}+\sum_{k\neq j}^\l \frac{\eps_j^2-\xi^2}{\eps_j^2-\eps_k^2}-2\sum_{i=1}^N\frac{\eps_j^2-\xi^2}{\eps_j^2-v_i^2}.
\end{equation*}
Then, rescale by $\dfrac{\eps_j^2}{\eps_j^2-\xi^2}$:
\begin{equation*}
\lambda_j^{(3)}=\left(\gamma+N-\frac{\l}{2}+\frac{1}{2}\right)\frac{\eps_j^2}{\eps_j^2-\xi^2}+\sum_{k\neq j}^\l \frac{\eps_j^2}{\eps_j^2-\eps_k^2}-2\sum_{i=1}^N\frac{\eps_j^2}{\eps_j^2-v_i^2}.
\end{equation*}
Choose $\gamma=-(\alpha+\beta+N-\l/2)$, which leads to
\begin{equation*}
\gamma+N-\frac{\l}{2}+\dfrac{1}{2}=-(\alpha+\beta)+\dfrac{1}{2}=-(\alpha+\beta+1)+\dfrac{3}{2}=\delta+\dfrac{3}{2}. 
\end{equation*}
Thus,
\begin{equation*}
\lambda_j^{(3)}=\left(\delta+\dfrac{3}{2}\right)\frac{\eps_j^2}{\eps_j^2-\xi^2}+\sum_{k\neq j}^\l \frac{\eps_j^2}{\eps_j^2-\eps_k^2}-2\sum_{i=1}^N\frac{\eps_j^2}{\eps_j^2-v_i^2}
\end{equation*}
is the same as \textbf{Rat. BQISM} (\ref{b_ev_rat*}) up to a constant term. Hence, \textbf{Trig. QISM} is equivalent to \textbf{Rat. BQISM} in the quasi-classical limit also on the level of the eigenvalue formula.

The difference of the constants in the eigenvalues
\begin{equation*}
\lambda_j^{(3)}-\eps_j{\lambda}_j^{rat}=\frac{3}{2}\frac{\eps_j^2}{\eps_j^2-\xi^2}-\frac{3}{4}=\frac{3}{4}\frac{\eps_j^2+\xi^2}{\eps_j^2-\xi^2}
\end{equation*}
is the same as the action of the difference of the conserved operators on the reference state:
\begin{equation*}
\tau_j^{(4)}\Omega-\eps_j{\tau}_j^{rat}\Omega
=\left(\frac{\eps_j^2}{\eps_j^2-\xi^2}\frac{1}{2}+\frac{\eps_j^2+\xi^2}{\eps_j^2-\xi^2}\frac{1}{2}-\frac{1}{4}\right)\Omega
= \left(\frac{3}{4}\frac{\eps_j^2+\xi^2}{\eps_j^2-\xi^2}\right)\Omega.
\end{equation*}

\ \\

\nid \underline{ {\bf Variable change $\# 2$}}

\ \\

\nid Here we want to transform the eigenvalue formula \textbf{Rat. BQISM} (\ref{b_ev_rat}) back into \textbf{Trig. BQISM} (\ref{b_ev_trig}). We start with \textbf{Rat. BQISM} in the form (\ref{b_ev_rat*}), multiplied by $\eps_j$:
\begin{equation*}
\tilde{\lambda}^{(1)}=\eps_j{\lambda}_j^{rat}=\,\frac{\delta\eps_j^2}{\eps_j^2-\xi^2}+\frac{3}{4}+\sum_{k\neq j}^\l\frac{\eps_j^2}{\eps_j^2-\eps_k^2}-
\sum_{i=1}^N\frac{2\eps_j^2}{\eps_j^2-v_i^2}.
\end{equation*}
We follow similar steps as in the case of the conserved operators, without the basis transformation. Start with the change of variables
\begin{equation*}
\eps_j\mapsto\frac{z_j-z_j^{-1}}{2},\ \ v_i\mapsto\frac{y_i-y_i^{-1}}{2}, \ \ \xi\mapsto\frac{\chi-\chi^{-1}}{2}.
\end{equation*}
This gives
\begin{equation*}
\tilde{\lambda}^{(2)}=\,\frac{\delta(z_j-z_j^{-1})^2}{(z_j-z_j^{-1})^2-(\chi-\chi^{-1})^2}+\frac{3}{4}+\sum_{k\neq j}^\l\frac{(z_j-z_j^{-1})^2}{(z_j-z_j^{-1})^2-(z_k-z_k^{-1})^2}-
\sum_{i=1}^N\frac{2(z_j-z_j^{-1})^2}{(z_j-z_j^{-1})^2-(y_i-y_i^{-1})^2}.
\end{equation*}
Now rescale by $\dfrac{z_j^2-z_j^{-2}}{(z_j-z_j^{-1})^2}$:
\begin{equation*}
\begin{aligned}
\tilde{\lambda}^{(3)}=&\,\frac{\delta(z_j^2-z_j^{-2})}{(z_j-z_j^{-1})^2-(\chi-\chi^{-1})^2}+\frac{3}{4}\frac{z_j^2-z_j^{-2}}{(z_j-z_j^{-1})^2}
+\sum_{k\neq j}^\l\frac{z_j^2-z_j^{-2}}{(z_j-z_j^{-1})^2-(z_k-z_k^{-1})^2}-\\
-&\sum_{i=1}^N\frac{2(z_j^2-z_j^{-2})}{(z_j-z_j^{-1})^2-(y_i-y_i^{-1})^2}.
\end{aligned}
\end{equation*}
Using the identity 
\begin{equation*}
\frac{(z_j^2-z_j^{-2})}{(z_j-z_j^{-1})^2-(z_k-z_k^{-1})^2}=\frac{1}{2}\left(\frac{z_j^2+z_k^2}{z_j^2-z_k^2}+\frac{z_j^2 z_k^2+1}{z_j^2 z_k^2-1}\right)
\end{equation*}
(and similar identities) we obtain
\begin{equation*}
\begin{aligned}
\tilde{\lambda}^{(3)}=&\,\frac{\delta}{2}\left(\frac{z_j^2+\chi^2}{z_j^2-\chi^2}+\frac{z_j^2 \chi^2+1}{z_j^2 \chi^2-1}\right)+\frac{3}{4}\frac{z_j^2-z_j^{-2}}{(z_j-z_j^{-1})^2}
+\frac{1}{2}\sum_{k\neq j}^\l\left(\frac{z_j^2+z_k^2}{z_j^2-z_k^2}+\frac{z_j^2 z_k^2+1}{z_j^2 z_k^2-1}\right)-\\
-&\sum_{i=1}^N\left(\frac{z_j^2+y_i^2}{z_j^2-y_i^2}+\frac{z_j^2 y_i^2+1}{z_j^2 y_i^2-1}\right).
\end{aligned}
\end{equation*}
This is the same, up to a constant term, as \textbf{Trig. BQISM} (\ref{b_ev_trig}) with the variable change $\eps_j=\ln z_j,\ v_i=\ln y_i,\ \xi=\ln\chi$:
\begin{equation*}
\begin{aligned}
{\lambda}^{trig}=&\,\frac{\delta}{2}\left(\frac{z_j^2+\chi^2}{z_j^2-\chi^2}+\frac{z_j^2 \chi^2+1}{z_j^2 \chi^2-1}\right)+\frac{3}{2}\frac{z_j^4+1}{z_j^4-1}
+\frac{1}{2}\sum_{k\neq j}^\l\left(\frac{z_j^2+z_k^2}{z_j^2-z_k^2}+\frac{z_j^2 z_k^2+1}{z_j^2 z_k^2-1}\right)-\\
-&\sum_{i=1}^N\left(\frac{z_j^2+y_i^2}{z_j^2-y_i^2}+\frac{z_j^2 y_i^2+1}{z_j^2 y_i^2-1}\right).
\end{aligned}
\end{equation*}
We have
\begin{equation}
\tilde{\lambda}^{(3)}-{\lambda}^{trig}=\frac{3}{4}\frac{z_j^2-z_j^{-2}}{(z_j-z_j^{-1})^2}-\frac{3}{2}\frac{z_j^4+1}{z_j^4-1}.
\label{evaldiff1}
\end{equation}

To check that the constants match with the constants from the conserved operators we need to compare
the expression (\ref{evaldiff1}) above with the action of $\tau_j^{(4)}-{\tau}_j^{trig}$ on
$\Omega$:
\begin{equation}\label{evaldiff2}
\begin{aligned}
\tau_j^{(4)}\Omega-{\tau}_j^{trig}\Omega&=\Bigg( \frac{1}{4}\frac{z_j^2-z_j^{-2}}{(z_j-z_j^{-1})^2}-
\frac{1}{2}\frac{(z_j-z_j^{-1})^2+(\chi-\chi^{-1})^2}{(z_j-z_j^{-1})^2}\frac{\chi^2(z_j^4-1)}{(z_j^2-\chi^2)(z_j^2
\chi^2-1)}\,-\\
&-\frac{1}{2}\frac{z_j^4+1}{z_j^4-1}+\frac{z_j^2}{z_j^4-1}\left(\frac{z_j^2-\chi^2}{\chi^2
z_j^2-1}+\frac{\chi^2 z_j^2-1}{z_j^2-\chi^2}\right) \Bigg)\Omega.
\end{aligned}
\end{equation}
The two expressions (\ref{evaldiff1}) and (\ref{evaldiff2}) are equivalent provided the following identity holds:
\begin{equation}\label{identtt}
\frac{z_j^4+1}{z_j^4-1}-\frac{1}{2}\frac{z_j^2-z_j^{-2}}{(z_j-z_j^{-1})^2}=
\frac{1}{2}\frac{(z_j-z_j^{-1})^2+
(\chi-\chi^{-1})^2}{(z_j-z_j^{-1})^2}\frac{\chi^2(z_j^4-1)}{(z_j^2-\chi^2)(z_j^2\chi^2-1)}
-\frac{z_j^2}{z_j^4-1}\left(\frac{z_j^2-\chi^2}{\chi^2 z_j^2-1}+\frac{\chi^2 z_j^2-1}{z_j^2-\chi^2}\right).
\end{equation}
Simplifying the left hand side of (\ref{identtt}) we find
\begin{equation*}
\frac{z_j^4+1}{z_j^4-1}-\frac{1}{2}\frac{z_j^2-z_j^{-2}}{(z_j-z_j^{-1})^2}=\frac{1}{2}\frac{z_j-z_j^{-1}}{z_j+z_j^{-1}}.
\end{equation*}
Modifying the right hand side of (\ref{identtt}) yields
\begin{equation*}
\begin{aligned}
&\frac{1}{2}\frac{(z_j-z_j^{-1})^2+(\chi-\chi^{-1})^2}{(z_j-z_j^{-1})^2}\frac{\chi^2(z_j^4-1)}{(z_j^2-\chi^2)(z_j^2\chi^2-1)}-
\frac{z_j^2}{z_j^4-1}\left(\frac{z_j^2-\chi^2}{\chi^2 z_j^2-1}+\frac{\chi^2 z_j^2-1}{z_j^2-\chi^2}\right)=\\
=&\,\frac{(z_j^2+z_j^{-2}+2)\chi^2(z_j^2-1)^2+(z_j^2+z_j^{-2}+2)z_j^2(\chi^2-1)^2-2(z_j^2-\chi^2)^2-2(z_j^2\chi^2-1)^2}{2(z_j-z_j^{-1})(z_j+z_j^{-1})(z_j^2-\chi^2)(z_j^2\chi^2-1)}=\\
=&\,\frac{(z_j-z_j^{-1})^2(z_j^2-\chi^2)(z_j^2\chi^2-1)}{2(z_j-z_j^{-1})(z_j+z_j^{-1})(z_j^2-\chi^2)(z_j^2\chi^2-1)}=\frac{1}{2}\frac{z_j-z_j^{-1}}{z_j+z_j^{-1}},
\end{aligned}
\end{equation*}
verifying that (\ref{identtt}) holds.


\ \\

\nid \underline{ {\bf Variable change $\# 3$} }

\ \\

\nid The variable change 3 is obtained in the same way as for the BAE and conserved operators, described in Sections \ref{sec:bae} and \ref{sec:im}.

\ \\

\nid \underline{ {\bf Reduction to the rational, twisted-periodic case} }

\ \\

\nid The rational limit of \textbf{Trig. QISM} (\ref{ev_trig}) gives
\begin{equation*}\label{ev_rat}
\lambda_j^{a.rat}=\gamma+\frac{1}{2}\sum_{k\neq j}^\l\frac{1}{\eps_j-\eps_k}-\sum_{i=1}^N\frac{1}{\eps_j-v_i}.
\end{equation*}
The rational limit of \textbf{Trig. BQISM$^\prime$} gives \textbf{Rat. BQISM$^\prime$}:
\begin{equation*}\label{b_ev_rat'}
\begin{aligned}
{\lambda}_j^{rat^\prime}=&\,\frac{\delta(\eps_j+\rho/2)}{(\eps_j+\rho/2)^2-\xi^2}+\frac{3}{2}\frac{1}{(2\eps_j+\rho)}+
\sum_{k\neq j}^\l\frac{\eps_j+\rho/2}{(\eps_j+\rho/2)^2-(\eps_k+\rho/2)^2}\,-\\
-&\sum_{i=1}^N\frac{2(\eps_j+\rho/2)}{(\eps_j+\rho/2)^2-(v_i+\rho/2)^2}.
\end{aligned}
\end{equation*}
Choose $\delta=\rho\gamma/2$. Then we see that, as $\rho\rightarrow\infty$,
${\lambda}_j^{rat^\prime}\rightarrow\lambda_j^{a.trig}$.



\begin{thebibliography}{99}
\bibitem{ado01} Amico L., Di Lorenzo, A., Osterloh A.: Integrable model for interacting electrons in metallic grains. Phys. Rev. Lett. {\bf 86}, 5759--5762 (2001)
\bibitem{aff01} Amico, L., Falci, G., Fazio, R.: The BCS model and the off-shell Bethe Ansatz for vertex models. J. Phys. A: Math. Gen. {\bf 34}, 6425--6434 (2001)
\bibitem{aacdfr03} Arnaudon, D., Avan, J., Crampe, N., Doikou, A., Frappat, L., Ragoucy, E.: Classification of reflection matrices related to (super-)Yangians and applications to open spin chain models. Nucl. Phys. B {\bf 668}, 469--505 (2003)
\bibitem{b72} Baxter, R.J.: Partition function of the eight-vertex lattice model. Ann. Phys. {\bf 70}, 193--228 (1972)
\bibitem{bd82} Belavin, A.A., Drinfel'd, V.G.: Solutions of the classical Yang--Baxter equation for simple Lie algebras. Funct. Anal. Appl. {\bf 16}, 159--180 (1982)
\bibitem{b31} Bethe, H.: Zur Theorie der Metalle. I. Eigenwerte und Eigenfunktionen der linearen Atomkette. Z. Phys. {\bf 71}, 205--226 (1931)
\bibitem{bc13} Belliard, S., Crampe, N.: Heisenberg $XXX$ model with general boundaries: eigenvectors from algebraic Bethe Ansatz. SIGMA {\bf 9}, 072 (2013)
\bibitem{crs97} Cambiaggio, M.C., Rivas, A.M.F., Saraceno, M.: (1997) Integrability of the pairing Hamiltonian. Nucl. Phys. A {\bf 624}, 157--167 (1997) 
\bibitem{cysw13} Cao, J., Yang, W.-L., Shi, K., Wang, Y.: Off-diagonal Bethe Ansatz solution of the $XXX$ spin chain with arbitrary boundary conditions. Nucl. Phys. B {\bf 875}, 152--165 (2013)
\bibitem{c84} Cherednik I.V.: Factorizing particles on a half-line and roots systems. Theoret. and Math. Phys. {\bf 61}, 977--983 (1984)
\bibitem{cmn13} Cirilo Ant\'onio, N., Manojlovi\'c, N., Nagy, Z.: Trigonometric $sl(2)$ Gaudin model with boundary terms. Rev. Math. Phys. {\bf 25}, 1343004 (2013)
\bibitem{dahog02} Di Lorenzo, A., Amico, L., Hikami, K., Osterloh, A., Giaquinta G.: Quasi-classical descendants of disordered vertex models with boundaries. Nucl. Phys. B {\bf 644}, 409--432 (2002)
\bibitem{dlr13} de Gier, J., Lee, A., Rasmussen, J.: Discrete holomorphicity and integrability in loop models with open boundaries. J. Stat. Mech.: Theor. Exp. P02029 (2013)
\bibitem{des01} Dukelsky, J., Esebbag, C., Schuck, P.: Class of exactly solvable pairing models. Phys. Rev. Lett. {\bf 87}, 066403 (2001)
\bibitem{dilsz10} Dunning, C., Iba\~nez, M., Links, J., Sierra, G., Zhao S.-Y.: Exact solution of the $p+ip$ pairing Hamiltoni9an and a hierarchy of integrable models. J. Stat. Mech.: Theor. Exp. P08025 (2010) 
\bibitem{fk10} Filali, G., Kitanine, N.: The partition function of the trigonometric $SOS$ model with a reflecting end. J. Stat. Mech. : Theor. Exp. L06001 (2010)
\bibitem{fk93} Foerster, A., Karowski, M.: The supersymmetric $t-J$ model with quantum group invariance. Nucl. Phys. B {\bf 408}, 512--534 (1993)
\bibitem{fs99} Frahm, H., Slavnov, N.A.: New solutions to the reflection equation and the projecting method. J. Phys. A: Math. Gen. {\bf 32}, 1547--1555 (1999) 
\bibitem{g08} Galleas, W.: Functional relations from the Yang-Baxter algebra: Eigenvalues of the $XXZ$ model with non-diagonal twisted and open bpoundary conditions. Nucl. Phys. B {\bf 790}, 524--542 (2008)
\bibitem{g76} Gaudin, M.: Diagonalisation d'une classe d'Hamiltoniens de spin. J. Phys. (Paris) {\bf 37}, 1087--1098 (1976)
\bibitem{h95} Hikami, K.: Gaudin magnet with boundary and generalized Knizhnik-Zamolodchikov equation. J. Phys. A: Math. Gen. {\bf 28}, 4997-5007 (1995)
\bibitem{ik14} Isaev, A.P., Kirillov, A.N.: Bethe subalgebras in Hecke algebra and Gaudin models. Lett. Math. Phys. {\bf 104}, 179--193 (2014)
\bibitem{iwwz13} Ikhlef, Y., Weston, R., Wheeler, M., Zinn-Justin, P.: Discrete holomorphicity and quantized affine algebras. J. Phys. A: Math. Theor. {\bf 46}, 265205 (2013)
\bibitem{ilsz09} Iba\~nez, M., Links, J., Sierra, G., Zhao S.-Y.: Exactly solvable pairing model for superconductors with $p_{x}+ip_{y}$-wave symmetry, Phys. Rev. B {\bf 79}, 180501 (2009)
\bibitem{kz94} Karowski, M., Zapletal, A.: Quantum-group-invariant integrable $n$-state vertex models with periodic boundary conditions. Nucl. Phys. B {\bf 419}, 567--588 (1994)
\bibitem{km03} Kulish, P.P., Manojlovi\'c, N.: Trigonometric $osp(1|2)$ Gaudin model. J. Math. Phys. {\bf 44}, 676--700 (2003)
\bibitem{ks79} Kulish, P.P., Sklyanin, E.K.: Quantum Inverse Scattering Method and the Heisenberg ferromagnet. Phys. Lett. {\bf 70A}, 461--4563 (1979)
\bibitem{ks92} Kulish, P.P., Sklyanin, E.K.: Algebraic structures related to reflection equations. J. Phys. A: Math. Gen. {\bf 25}, 5963--5975 (1992)
\bibitem{ml06} Malara, R., Lima-Santos, A.: On ${\mathcal A}^{(1)}_{n-1}$, ${\mathcal B}^{(1)}_n$, 
${\mathcal C}^{(1)}_n$, ${\mathcal D}^{(1)}_n$, ${\mathcal A}^{(2)}_{2n}$, ${\mathcal A}^{(2)}_{2n+1}$, and 
${\mathcal D}^{(2)}_{n+1}$ reflection $K$-matrices. J. Stat. Mech.: Theor. Exp. P09013 (2006) 
\bibitem{m64} McGuire, J.B.: Study of exactly soluble one-dimensional $N$-body problems. J. Math. Phys. {\bf 5}, 622--636 (1964)
\bibitem{mrm05} Melo, C.S., Ribeiro, G.A.P., Martins, M.J.: Bethe ansatz for the $XXX-S$ chain with non-diagonal open boundaries. Nucl. Phys. B {\bf 711}, 565--603 (2005)
\bibitem{mn92} Mezincescu, L., Nepomechie, R.I.: Analytical Bethe Ansatz for quantum-algebra-invariant spin chains. Nucl. Phys. B {\bf 372}, 597-621 (1992)
\bibitem{n12} Niccoli, G.: Non-diagonal open spin-1/2 $XXZ$ quantum chains by separation of variables: complete spectrum and matrix elements of some quasi-local operators. J. Stat. Mech.: Theor. Exp. P100025 (2012)
\bibitem{o03} Ovchinnikov, A.A.: Exactly solvable discrete BCS-type Hamiltonians and the six-vertex model. Nucl. Phys. B {\bf 703},  363--390 (2003)
\bibitem{r63} Richardson, R.W.: A restricted class of exact eigenstates of the pairing-force Hamiltonian, {\it  Phys. Lett.} {\bf 3} 277--279 (1963)
\bibitem{rdo10} Rombouts, M.A.S., Dukelsky, J., Ortiz, G.: Quantum phase diagram of the integrable $p_{x}+ip_{y}$ fermionic superfluid. Phys. Rev. B {\bf 82}, 224510 (2010)
\bibitem{s00} Sierra, G.: Conformal field theory and the exact solution of the BCS Hamiltonian. Nucl. Phys. B {\bf 572}, 517--534 (2000)
\bibitem{s87} Sklyanin, E.K.: Boundary conditions for integrable equations. Funct. Anal. Appl. {\bf 21}, 164--166 (1987)
\bibitem{s88} Sklyanin, E.K.: Boundary conditions for integrable quantum systems. J. Phys. A: Math. Gen. {\bf 21}, 2375--2389 (1988)
\bibitem{s89} Sklyanin, E.K.: Separation of variables in the Gaudin model. J. Sov. Math. {\bf 47}, 2473--2488  (1989)
\bibitem{s07} Skrypnyk, T.: Generalized Gaudin spin chains, nonskew symmetric $r$-matrices, and reflection equation algebras. J. Math. Phys. {\bf 48}, 113521 (2007)
\bibitem{s09} Skrypnyk, T.: Non-skew-symmetric classical $r$-matrices and integrable cases of the reduced BCS model. J. Phys. A: Math. Theor. {\bf 42}, 472004 (2009) 
\bibitem{s10} Skrypnyk, T.: Generalized Gaudin systems in an external magnetic field and reflection equation algebras. J. Stat. Mech.: Theor. Exp. P06028 (2010)
\bibitem{s12} Skrypnyk, T.: Rational $r$-matrices, higher rank Lie algebra and integrable proton-neutron BCS models. Nucl. Phys. B {\bf 863}, 435--469 (2012)
\bibitem{s13} Skrypnyk, T.: ``${\mathbb Z}_2$-graded '' Gaudin models and analytical Bethe ansatz. Nucl. Phys. B {\bf 870}, 495--529 (2013)
\bibitem{tf79} Takhtadzhan, L.A., Faddeev, L.D.: The quantum method of the inverse problem and the Heisenberg $XYZ$ model. Russ. Math. Surv. {\bf 34}, 11--68 (1979)
\bibitem{vdp02} von Delft, J., Poghossian, R.: Algebraic Bethe Ansatz for a discrete-state BCS pairing model. Phys. Rev. B {\bf 66} 134502 (2002)
\bibitem{vdr01} von Delft, J., Ralph, D.C.: Spectroscopy of discrete energy levels in ultrasmall metallic grains. Phys. Rep. {\bf 345}, 61--173  (2001)
\bibitem{y67} Yang, C.N.: Some exact results for the many-body problem in one dimension with repulsive delta-function interaction. Phys. Rev. Lett. {\bf 19}, 1312--1315  (1967)
\bibitem{yzg04} Yang, W.-L., Zhang, Y.-Z., Gould, M.D.: Exact solution of the $XXZ$ Gaudin model with generic open boundaries. Nucl. Phys. B {\bf 698}, 503--516 (2004)
\bibitem{zlmg02} Zhou, H.-Q., Links, J., McKenzie, R.H., Gould, M.D.: Superconducting correlations in metallic nanograins: exact solution of the BCS model by the algebraic Bethe Ansatz. Phys. Rev. B {\bf 65},  060502(R) (2002)
\end{thebibliography}

\end{document}